\documentclass[reprint,amsmath,amssymb,aps,superscriptaddress,nofootinbib]{revtex4-2}

\usepackage{tikz}

\usepackage{dcolumn}
\usepackage{bm}
\usepackage[english]{babel}
\usepackage{amsmath, amssymb}
\usepackage[utf8]{inputenc}
\usepackage[T1]{fontenc}
\usepackage{graphicx}
\usepackage{booktabs}
\usepackage{enumerate}
\usepackage{multirow}
\usepackage{subfigure}
\usepackage{dsfont}
\usepackage{slashed}
\usepackage{textcomp}
\usepackage{url}
\usepackage{float}
\usepackage{hyperref}
\usepackage[figure]{hypcap}
\usepackage{physics}
\usepackage[space-before-unit, detect-weight, detect-display-math, detect-inline-weight=math, per-mode=fraction]{siunitx}
\usepackage{pgf}
\usepackage{color}
\usepackage{changepage}
\usepackage{import} 

\newcommand\inputpgf[2]{{
\let\pgfimageWithoutPath\pgfimage
\renewcommand{\pgfimage}[2][]{\pgfimageWithoutPath[##1]{#1/##2}}
\input{#1/#2}
}}

\begin{document}

\title{Combined sensitivity to the neutrino mass ordering\\ with JUNO, the IceCube Upgrade, and PINGU}

\affiliation{III. Physikalisches Institut, RWTH Aachen University, D-52056 Aachen, Germany}
\affiliation{Department of Physics, University of Adelaide, Adelaide, 5005, Australia}
\affiliation{Dept. of Physics and Astronomy, University of Alaska Anchorage, 3211 Providence Dr., Anchorage, AK 99508, USA}
\affiliation{Dept. of Physics, University of Texas at Arlington, 502 Yates St., Science Hall Rm 108, Box 19059, Arlington, TX 76019, USA}
\affiliation{CTSPS, Clark-Atlanta University, Atlanta, GA 30314, USA}
\affiliation{School of Physics and Center for Relativistic Astrophysics, Georgia Institute of Technology, Atlanta, GA 30332, USA}
\affiliation{Dept. of Physics, Southern University, Baton Rouge, LA 70813, USA}
\affiliation{Dept. of Physics, University of California, Berkeley, CA 94720, USA}
\affiliation{Lawrence Berkeley National Laboratory, Berkeley, CA 94720, USA}
\affiliation{Institut f{\"u}r Physik, Humboldt-Universit{\"a}t zu Berlin, D-12489 Berlin, Germany}
\affiliation{Fakult{\"a}t f{\"u}r Physik {\&} Astronomie, Ruhr-Universit{\"a}t Bochum, D-44780 Bochum, Germany}
\affiliation{Universit{\'e} Libre de Bruxelles, Science Faculty CP230, B-1050 Brussels, Belgium}
\affiliation{Vrije Universiteit Brussel (VUB), Dienst ELEM, B-1050 Brussels, Belgium}
\affiliation{Dept. of Physics, Massachusetts Institute of Technology, Cambridge, MA 02139, USA}
\affiliation{Dept. of Physics and Institute for Global Prominent Research, Chiba University, Chiba 263-8522, Japan}
\affiliation{Dept. of Physics and Astronomy, University of Canterbury, Private Bag 4800, Christchurch, New Zealand}
\affiliation{Dept. of Physics, University of Maryland, College Park, MD 20742, USA}
\affiliation{Dept. of Astronomy, Ohio State University, Columbus, OH 43210, USA}
\affiliation{Dept. of Physics and Center for Cosmology and Astro-Particle Physics, Ohio State University, Columbus, OH 43210, USA}
\affiliation{Niels Bohr Institute, University of Copenhagen, DK-2100 Copenhagen, Denmark}
\affiliation{Dept. of Physics, TU Dortmund University, D-44221 Dortmund, Germany}
\affiliation{Dept. of Physics and Astronomy, Michigan State University, East Lansing, MI 48824, USA}
\affiliation{Dept. of Physics, University of Alberta, Edmonton, Alberta, Canada T6G 2E1}
\affiliation{Erlangen Centre for Astroparticle Physics, Friedrich-Alexander-Universit{\"a}t Erlangen-N{\"u}rnberg, D-91058 Erlangen, Germany}
\affiliation{Physik-department, Technische Universit{\"a}t M{\"u}nchen, D-85748 Garching, Germany}
\affiliation{D{\'e}partement de physique nucl{\'e}aire et corpusculaire, Universit{\'e} de Gen{\`e}ve, CH-1211 Gen{\`e}ve, Switzerland}
\affiliation{Dept. of Physics and Astronomy, University of Gent, B-9000 Gent, Belgium}
\affiliation{Dept. of Physics and Astronomy, University of California, Irvine, CA 92697, USA}
\affiliation{Karlsruhe Institute of Technology, Institut f{\"u}r Kernphysik, D-76021 Karlsruhe, Germany}
\affiliation{Dept. of Physics and Astronomy, University of Kansas, Lawrence, KS 66045, USA}
\affiliation{SNOLAB, 1039 Regional Road 24, Creighton Mine 9, Lively, ON, Canada P3Y 1N2}
\affiliation{School of Physics and Astronomy, Queen Mary University of London, London E1 4NS, United Kingdom}
\affiliation{Department of Physics and Astronomy, UCLA, Los Angeles, CA 90095, USA}
\affiliation{Department of Physics, Mercer University, Macon, GA 31207-0001, USA}
\affiliation{Dept. of Astronomy, University of Wisconsin, Madison, WI 53706, USA}
\affiliation{Dept. of Physics and Wisconsin IceCube Particle Astrophysics Center, University of Wisconsin, Madison, WI 53706, USA}
\affiliation{Institute of Physics, University of Mainz, Staudinger Weg 7, D-55099 Mainz, Germany}
\affiliation{School of Physics and Astronomy, The University of Manchester, Oxford Road, Manchester, M13 9PL, United Kingdom}
\affiliation{Department of Physics, Marquette University, Milwaukee, WI, 53201, USA}
\affiliation{Institut f{\"u}r Kernphysik, Westf{\"a}lische Wilhelms-Universit{\"a}t M{\"u}nster, D-48149 M{\"u}nster, Germany}
\affiliation{Bartol Research Institute and Dept. of Physics and Astronomy, University of Delaware, Newark, DE 19716, USA}
\affiliation{Dept. of Physics, Yale University, New Haven, CT 06520, USA}
\affiliation{Columbia Astrophysics and Nevis Laboratories, Columbia University, New York, NY 10027, USA}
\affiliation{Dept. of Physics, University of Notre Dame du Lac, 225 Nieuwland Science Hall, Notre Dame, IN 46556-5670, USA}
\affiliation{Dept. of Physics, University of Oxford, Parks Road, Oxford OX1 3PU, UK}
\affiliation{Dept. of Physics, Drexel University, 3141 Chestnut Street, Philadelphia, PA 19104, USA}
\affiliation{Physics Department, South Dakota School of Mines and Technology, Rapid City, SD 57701, USA}
\affiliation{Dept. of Physics, University of Wisconsin, River Falls, WI 54022, USA}
\affiliation{Dept. of Physics and Astronomy, University of Rochester, Rochester, NY 14627, USA}
\affiliation{Oskar Klein Centre and Dept. of Physics, Stockholm University, SE-10691 Stockholm, Sweden}
\affiliation{Dept. of Physics and Astronomy, Stony Brook University, Stony Brook, NY 11794-3800, USA}
\affiliation{Dept. of Physics, Sungkyunkwan University, Suwon 16419, Korea}
\affiliation{Institute of Basic Science, Sungkyunkwan University, Suwon 16419, Korea}
\affiliation{Earthquake Research Institute, University of Tokyo, Bunkyo, Tokyo 113-0032, Japan}
\affiliation{Dept. of Physics and Astronomy, University of Alabama, Tuscaloosa, AL 35487, USA}
\affiliation{Dept. of Astronomy and Astrophysics, Pennsylvania State University, University Park, PA 16802, USA}
\affiliation{Dept. of Physics, Pennsylvania State University, University Park, PA 16802, USA}
\affiliation{Dept. of Physics and Astronomy, Uppsala University, Box 516, S-75120 Uppsala, Sweden}
\affiliation{Dept. of Physics, University of Wuppertal, D-42119 Wuppertal, Germany}
\affiliation{DESY, D-15738 Zeuthen, Germany}
\affiliation{Beijing Institute of Spacecraft Environment Engineering, Beijing 100094, China}
\affiliation{Sun Yat-Sen University, Guangzhou, China}
\affiliation{Institute of Experimental Physics, University of Hamburg, Hamburg, Germany}
\affiliation{Institut f\"ur Kernphysik 2, Forschungszentrum J\"ulich, 52425 J\"ulich, Germany}
\affiliation{Department of Physics, University of Jyv\"askyl\"a, Jyv\"askyl\"a, Finland}
\affiliation{Gran Sasso Science Institute, 67100 L'Aquila, Italy}
\affiliation{Dipartimento di Fisica Università and INFN Milano, 20133 Milano, Italy}
\affiliation{Institute of Nuclear Research, Russian Academy of Sciences, Moscow, 117312, Russia}
\affiliation{Faculty of Physics, Lomonosov Moscow State University, Moscow, 119991, Russia}
\affiliation{SUBATECH, CNRS/IN2P3, Université de Nantes, IMT-Atlantique, 44307 Nantes, France}
\affiliation{Laboratoire de l'accélérateur linéaire Orsay, 91440 Orsay, France}
\affiliation{Dipartimento di Fisica e Astronomia dell'Università di Padova and INFN Sezione di Padova, Padova, Italy}
\affiliation{Astro-Particle Physics Laboratory, CNRS/CEA/Paris7/Observatoire de Paris, Paris, France}
\affiliation{University of Roma Tre and INFN Sezione Roma Tre, Roma, Italy}
\affiliation{Eberhard Karls Universit\"at T\"ubingen, Physikalisches Institut, T\"ubingen, Germany}

\author{M. G. Aartsen}
\affiliation{Dept. of Physics and Astronomy, University of Canterbury, Private Bag 4800, Christchurch, New Zealand}
\author{M. Ackermann}
\affiliation{DESY, D-15738 Zeuthen, Germany}
\author{J. Adams}
\affiliation{Dept. of Physics and Astronomy, University of Canterbury, Private Bag 4800, Christchurch, New Zealand}
\author{J. A. Aguilar}
\affiliation{Universit{\'e} Libre de Bruxelles, Science Faculty CP230, B-1050 Brussels, Belgium}
\author{M. Ahlers}
\affiliation{Niels Bohr Institute, University of Copenhagen, DK-2100 Copenhagen, Denmark}
\author{M. Ahrens}
\affiliation{Oskar Klein Centre and Dept. of Physics, Stockholm University, SE-10691 Stockholm, Sweden}
\author{C. Alispach}
\affiliation{D{\'e}partement de physique nucl{\'e}aire et corpusculaire, Universit{\'e} de Gen{\`e}ve, CH-1211 Gen{\`e}ve, Switzerland}
\author{K. Andeen}
\affiliation{Department of Physics, Marquette University, Milwaukee, WI, 53201, USA}
\author{T. Anderson}
\affiliation{Dept. of Physics, Pennsylvania State University, University Park, PA 16802, USA}
\author{I. Ansseau}
\affiliation{Universit{\'e} Libre de Bruxelles, Science Faculty CP230, B-1050 Brussels, Belgium}
% ...

\collaboration{IceCube-Gen2 Collaboration}
 \email{analysis@icecube.wisc.edu}
\noaffiliation

\author{T.~J.~C. Bezerra}
\affiliation{SUBATECH, CNRS/IN2P3, Université de Nantes, IMT-Atlantique, 44307 Nantes, France}
\author{T. Birkenfeld}
\affiliation{III. Physikalisches Institut, RWTH Aachen University, D-52056 Aachen, Germany}
\author{D. Blum}
\affiliation{Eberhard Karls Universit\"at T\"ubingen, Physikalisches Institut, T\"ubingen, Germany}
\author{M. Bongrand}
\affiliation{Laboratoire de l'accélérateur linéaire Orsay, 91440 Orsay, France}
\author{A. Cabrera}
\affiliation{Laboratoire de l'accélérateur linéaire Orsay, 91440 Orsay, France}
\author{Y. P. Cheng}
\affiliation{Beijing Institute of Spacecraft Environment Engineering, Beijing 100094, China}
\author{W. Depnering}
\affiliation{Institute of Physics, University of Mainz, Staudinger Weg 7, D-55099 Mainz, Germany}
\author{O. D\"otterl}
\affiliation{Physik-department, Technische Universit{\"a}t M{\"u}nchen, D-85748 Garching, Germany}
\author{T. Enqvist}
\affiliation{Department of Physics, University of Jyv\"askyl\"a, Jyv\"askyl\"a, Finland}
\author{H. Enzmann}
\affiliation{Institute of Physics, University of Mainz, Staudinger Weg 7, D-55099 Mainz, Germany}
\author{C. Genster}
\affiliation{III. Physikalisches Institut, RWTH Aachen University, D-52056 Aachen, Germany}
\affiliation{Institut f\"ur Kernphysik 2, Forschungszentrum J\"ulich, 52425 J\"ulich, Germany}
\author{M. Grassi}
\affiliation{Laboratoire de l'accélérateur linéaire Orsay, 91440 Orsay, France}
\author{A. S. G\"ottel}
\affiliation{III. Physikalisches Institut, RWTH Aachen University, D-52056 Aachen, Germany}
\affiliation{Institut f\"ur Kernphysik 2, Forschungszentrum J\"ulich, 52425 J\"ulich, Germany}
\author{P. Hackspacher}
\affiliation{Institute of Physics, University of Mainz, Staudinger Weg 7, D-55099 Mainz, Germany}
\author{C. Hagner}
\affiliation{Institute of Experimental Physics, University of Hamburg, Hamburg, Germany}
\author{Y. Han}
\affiliation{Laboratoire de l'accélérateur linéaire Orsay, 91440 Orsay, France}
\author{T. Heinz}
\affiliation{Eberhard Karls Universit\"at T\"ubingen, Physikalisches Institut, T\"ubingen, Germany}
\author{P. Kampmann}
\affiliation{III. Physikalisches Institut, RWTH Aachen University, D-52056 Aachen, Germany}
\affiliation{Institut f\"ur Kernphysik 2, Forschungszentrum J\"ulich, 52425 J\"ulich, Germany}
\author{P. Kuusiniemi}
\affiliation{Department of Physics, University of Jyv\"askyl\"a, Jyv\"askyl\"a, Finland}
\author{T. Lachenmaier}
\affiliation{Eberhard Karls Universit\"at T\"ubingen, Physikalisches Institut, T\"ubingen, Germany}
\author{K. Loo}
\affiliation{Institute of Physics, University of Mainz, Staudinger Weg 7, D-55099 Mainz, Germany}
\author{S. Lorenz}
\affiliation{Institute of Physics, University of Mainz, Staudinger Weg 7, D-55099 Mainz, Germany}
\author{B. Lubsandorzhiev}
\affiliation{Institute of Nuclear Research, Russian Academy of Sciences, Moscow, 117312, Russia}
\author{L. Ludhova}
\affiliation{III. Physikalisches Institut, RWTH Aachen University, D-52056 Aachen, Germany}
\affiliation{Institut f\"ur Kernphysik 2, Forschungszentrum J\"ulich, 52425 J\"ulich, Germany}
\author{Y. Malyshkin}
\affiliation{University of Roma Tre and INFN Sezione Roma Tre, Roma, Italy}
\author{D. Meyhöfer}
\affiliation{Institute of Experimental Physics, University of Hamburg, Hamburg, Germany}
\author{L. Miramonti}
\affiliation{Dipartimento di Fisica Università and INFN Milano, 20133 Milano, Italy}
\author{A. M\"uller}
\affiliation{Eberhard Karls Universit\"at T\"ubingen, Physikalisches Institut, T\"ubingen, Germany}
\author{L. J. N. Oberauer}
\affiliation{Physik-department, Technische Universit{\"a}t M{\"u}nchen, D-85748 Garching, Germany}
\author{O. Pilarczyk}
\affiliation{Institute of Physics, University of Mainz, Staudinger Weg 7, D-55099 Mainz, Germany}
\author{H. Rebber}
\affiliation{Institute of Experimental Physics, University of Hamburg, Hamburg, Germany}
\author{J. Sawatzki}
\affiliation{Physik-department, Technische Universit{\"a}t M{\"u}nchen, D-85748 Garching, Germany}
\author{M. Schever}
\affiliation{III. Physikalisches Institut, RWTH Aachen University, D-52056 Aachen, Germany}
\affiliation{Institut f\"ur Kernphysik 2, Forschungszentrum J\"ulich, 52425 J\"ulich, Germany}
\author{K. Schweizer}
\affiliation{Physik-department, Technische Universit{\"a}t M{\"u}nchen, D-85748 Garching, Germany}
\author{M. Settimo}
\affiliation{SUBATECH, CNRS/IN2P3, Université de Nantes, IMT-Atlantique, 44307 Nantes, France}
\author{C. Sirignano}
\affiliation{Dipartimento di Fisica e Astronomia dell'Università di Padova and INFN Sezione di Padova, Padova, Italy}
\author{M. Smirnov}
\affiliation{Sun Yat-Sen University, Guangzhou, China}
\author{A. Stahl}
\affiliation{III. Physikalisches Institut, RWTH Aachen University, D-52056 Aachen, Germany}
\author{H. T. J. Steiger}
\affiliation{Physik-department, Technische Universit{\"a}t M{\"u}nchen, D-85748 Garching, Germany}
\author{J. Steinmann}
\affiliation{III. Physikalisches Institut, RWTH Aachen University, D-52056 Aachen, Germany}
\author{T. Sterr}
\affiliation{Eberhard Karls Universit\"at T\"ubingen, Physikalisches Institut, T\"ubingen, Germany}
\author{M. R. Stock}
\affiliation{Physik-department, Technische Universit{\"a}t M{\"u}nchen, D-85748 Garching, Germany}
\author{A. Studenikin}
\affiliation{Faculty of Physics, Lomonosov Moscow State University, Moscow 119991, Russia}
\author{A. Tietzsch}
\affiliation{Eberhard Karls Universit\"at T\"ubingen, Physikalisches Institut, T\"ubingen, Germany}
\author{W. H. Trzaska}
\affiliation{Department of Physics, University of Jyv\"askyl\"a, Jyv\"askyl\"a, Finland}
\author{B. Viaud}
\affiliation{SUBATECH, CNRS/IN2P3, Université de Nantes, IMT-Atlantique, 44307 Nantes, France}
\author{C. Volpe}
\affiliation{Astro-Particle Physics Laboratory, CNRS/CEA/Paris7/Observatoire de Paris, Paris, France}
\author{W. Wang}
\affiliation{Sun Yat-Sen University, Guangzhou, China}
\author{B. S. Wonsak}
\affiliation{Institute of Experimental Physics, University of Hamburg, Hamburg, Germany}
\author{M. Wurm}
\affiliation{Institute of Physics, University of Mainz, Staudinger Weg 7, D-55099 Mainz, Germany}
\author{C. Wysotzki}
\affiliation{III. Physikalisches Institut, RWTH Aachen University, D-52056 Aachen, Germany}
\author{Y. Xu}
\affiliation{III. Physikalisches Institut, RWTH Aachen University, D-52056 Aachen, Germany}
\affiliation{Institut f\"ur Kernphysik 2, Forschungszentrum J\"ulich, 52425 J\"ulich, Germany}
\author{D. Xuefeng}
\affiliation{Gran Sasso Science Institute, 67100 L'Aquila, Italy}
\affiliation{Dipartimento di Fisica Università and INFN Milano, 20133 Milano, Italy}
\author{F. Yermia}
\affiliation{SUBATECH, CNRS/IN2P3, Université de Nantes, IMT-Atlantique, 44307 Nantes, France}

\collaboration{JUNO Collaboration Members}
 \email{juno_collaboration@juno.ihep.ac.cn}
\noaffiliation

\date{\today}

\begin{abstract}
  The ordering of the neutrino mass eigenstates is one of the fundamental open questions in neutrino physics. While current-generation neutrino oscillation experiments are able to produce moderate indications on this ordering, upcoming experiments of the next generation aim to provide conclusive evidence. In this paper we study the combined performance of the two future multi-purpose neutrino oscillation experiments JUNO and the IceCube Upgrade, which employ two very distinct and complementary routes towards the neutrino mass ordering. The approach pursued by the \SI{20}{\kilo\tonne} medium-baseline reactor neutrino experiment JUNO consists of a careful investigation of the energy spectrum of oscillated $\bar{\nu}_e$ produced by ten nuclear reactor cores. The IceCube Upgrade, on the other hand, which consists of seven additional densely instrumented strings deployed in the center of IceCube DeepCore, will observe large numbers of atmospheric neutrinos that have undergone oscillations affected by Earth matter. In a joint fit with both approaches, tension occurs between their preferred mass-squared differences $ \Delta m_{31}^{2}=m_{3}^{2}-m_{1}^{2} $ within the wrong mass ordering. In the case of JUNO and the IceCube Upgrade, this allows to exclude the wrong ordering at $>5\sigma$ on a timescale of 3--7 years --- even under circumstances that are unfavorable to the experiments' individual sensitivities. For PINGU, a 26-string detector array designed as a potential low-energy extension to IceCube, the inverted ordering could be excluded within 1.5 years (3 years for the normal ordering) in a joint analysis.
\end{abstract}

\maketitle

\section{Introduction}
\label{chap:motivation}

The neutrino mass ordering (NMO) is one of the most important questions to be solved in the field of neutrino oscillation physics. It impacts the limit on the (absolute) incoherent sum of neutrino masses from cosmology~\cite{cosmo_mass}, has important implications for the searches for neutrinoless double-$\beta$ decay~\cite{double_beta}, and leads to a better understanding of the (flavor-mass) mixing in the lepton sector~\cite{King:2014nza}. In addition, knowledge of the NMO is an important prerequisite for unambiguously measuring leptonic charge-parity (CP) violation~\cite{Barger:2001yr}.

By convention, $m_1$ is assumed to be lighter than $m_2$, and $\Delta m^2_{21}$ is taken to be the smallest mass-squared difference in magnitude~\cite{deGouvea:2008nm}. A global analysis of solar neutrino and KamLAND data yields $\Delta m^2_{21} \equiv m^2_2 - m^2_1 \approx+ \SI{7.5e-5}{\electronvolt\squared}$ (known to a precision of about \SI{2}{\percent})~\cite{PDG2018}, but leaves us with two possible realizations of the neutrino mass ordering. The ``normal'' ordering (NO) has $ m_{1}<m_{2}<m_{3} $, whereas the ``inverted'' ordering (IO) has $ m_{3}<m_{1}<m_{2} $. They are distinguished by the sign of the mass-squared difference $ \Delta m_{31}^{2} $ (or, equivalently, $ \Delta m_{32}^{2} $), with $\Delta m^2_{31} > 0$ in the case of the NO and $\Delta m^2_{31} < 0$ in the case of the IO.

The global analysis of atmospheric, accelerator, and medium-baseline reactor neutrino oscillation experiments determines the absolute value $|\Delta m^2_{31(32)}|\approx \SI{2.5e-3}{\electronvolt\squared}$ to better than \SI{2}{\percent} precision~\cite{PDG2018}. The sign of this larger of the two independent mass-squared differences still remains unknown; only recently have global fits begun to demonstrate a growing preference for the NO~\cite{Nufit4.0,nufit4.0url,deSalas:2017kay,Capozzi:2018ubv}.

The most sensitive current long-baseline accelerator experiments NO$\nu$A and T2K, as well as the atmospheric neutrino experiment Super-K, each show some preference for the NO~\cite{NOvA:2018gge, Abe:2018wpn, Abe:2017aap}, through Earth matter effects on $\Delta m^2_{31}$-driven oscillations~\cite{Wolfenstein:1978ue,Mikheev:1986wj,Petcov:1986qg,Akhmedov:2006hb}. Subtle synergy effects from medium-baseline reactor experiments without NMO sensitivity (Daya Bay, Double Chooz, RENO) significantly contribute to the global constraint, owing to tension in the preferred values of $\Delta m^2_{31}$ within the IO~\cite[Fig.~9]{Nufit4.0}. Whether a $5\sigma$ measurement of the NMO will be obtained with a combination of current experiments is unclear. This explains the need for additional experimental efforts~\cite{deSalas:2018bym}.

An approach that has so far not been realized consists of exploring subleading survival probability terms arising from the difference between $\Delta m^2_{31}$ and $\Delta m^2_{32}$ for electron anti-neutrinos detected at a distance of $\mathcal{O}(\SI{50}{\kilo\metre})$ from a nuclear reactor~\cite{Petcov:2001sy}. This technique, pursued by the medium-baseline reactor experiments JUNO~\cite{JUNO} and RENO-50~\cite{Kim:2014rfa}, provides a promising route forward on its own. In addition, it has been shown that its combination with an independent measurement of $|\Delta m^2_{31}|$ by a long-baseline experiment further enhances the NMO sensitivity~\cite{Li:2013zyd}. An even stronger enhancement can be expected from the combination with next-generation long-baseline oscillation data, collected by atmospheric neutrino oscillation experiments such as Hyper-K~\cite{Abe:2011ts,Abe:2018uyc}, ICAL@INO~\cite{Kumar:2017sdq}, ORCA~\cite{Adrian-Martinez:2016fdl} and PINGU~\cite{TheIceCube-Gen2:2016cap,PINGU}, or the accelerator experiments DUNE~\cite{Acciarri:2015uup} and T2HK/T2HKK~\cite{Abe:2015zbg,Abe:2016ero}. All of these will rely on more or less pronounced effects of Earth matter on the neutrino flavor evolution in order to determine the NMO.

In this paper we study in detail the expected NMO capabilities of the combination of the medium-baseline reactor experiment JUNO and either the IceCube Upgrade~\cite{upgrade_icrc:2019} (funded) or PINGU (proposed), which are low-energy extensions of the IceCube detector that are sensitive to atmospheric neutrino oscillations. The individual published projected NMO sensitivities of JUNO~\cite{JUNO} and PINGU~\cite{TheIceCube-Gen2:2016cap,PINGU} exceed those of the current-generation oscillation experiments. Even more crucially, as has been shown in~\cite{BlennowSchwetz}, the joint fit of the oscillation data from JUNO and PINGU will profit from strong synergy effects. These are brought about by tension between the fit values of $\Delta m^2_{31}$ within the wrong NMO---similar to but stronger in magnitude than the tension encountered by the current global fits referred to above. Since the underlying neutrino oscillation physics is the same, synergy will also occur between JUNO and the imminent IceCube Upgrade. We will demonstrate explicitly that even in this combination, the synergy is strong enough to decisively exclude the wrong ordering.

JUNO~\cite{JUNO} is a \SI{20}{\kilo\tonne} liquid scintillator (linear alkylbenzene) detector currently under design and construction near Jiangmen in South China. Deployed in an underground laboratory with \SI{700}{\metre} overburden, it is designed to measure the disappearance of \si{MeV}-energy $ \bar{\nu}_{e} $ produced by the Yangjiang and Taishan nuclear power plants at a distance of approximately \SI{53}{\kilo\metre}. At the start of data taking, anticipated in 2021, the number of operational reactor cores is expected to be eight instead of the baseline number of ten. Therefore, we conservatively perform the JUNO analysis with both configurations. In the detector, the reactor $\bar{\nu}_e$ convert to positrons via the inverse beta decay (IBD) process~\cite{Vogel:1999zy,Formaggio:2013kya} on protons. This results in a characteristic pair of prompt and delayed photon showers, which are detected by about \SI{17 000}{} 20-inch photomultiplier tubes (PMTs). The high number of photoelectrons per event (1200 p.e./MeV)~\cite{JUNO} allows for a percent-level neutrino energy resolution. As a result, the NMO can be determined through a precise measurement of the oscillatory fine structure imprinted on the neutrino energy spectrum.

The IceCube-Gen2 facility is the planned next-generation extension of IceCube. It will integrate the operating IceCube detector together with four new components: (1) the IceCube-Gen2 optical array complemented by (2) the low-energy core, (3) the radio array, and (4) the surface array~\cite{Aartsen:2019swn}.

An essential first step in the path towards IceCube-Gen2 is the IceCube Upgrade project~\cite{upgrade_icrc:2019} (funded and under construction, scheduled for deployment in the 2022/2023 polar season), in the following referred to simply as ``Upgrade'' for brevity. It provides the initial, highly sensitive low-energy-core array and serves simultaneously as a path-finder mission for the larger IceCube-Gen2 optical high-energy array. A more ambitious detector configuration than the Upgrade array is PINGU~\cite{TheIceCube-Gen2:2016cap,PINGU}, that would consist of 26 densely instrumented strings.
Both the Upgrade and PINGU will in-fill the existing IceCube DeepCore~\cite{Collaboration:2011ym} region in deep glacial ice, thereby increasing the neutrino detection efficiency for energies of a few \si{\giga\electronvolt}. In this paper we consider the sensitivity of both the Upgrade and the PINGU array.

Our simulation of the Upgrade assumes that its 7 strings are instrumented with optical modules identical in type to the high quantum-efficiency (HQE) DOMs deployed in DeepCore~\cite{Abbasi:2008aa,Abbasi:2010vc}. Each string holds 125 modules, spaced \SI{2.4}{\metre} apart. The spacing between the strings is about \SI{25}{\metre}. In the meantime, some aspects related to detector instrumentation and layout have received updates~\cite{upgrade_icrc:2019}. For example, the string spacing has been adapted for calibration purposes, and the majority of sensors will be optical modules of the next generation.

In the case of PINGU, we assume that all 26 strings are composed of 192 HQE DOMs. Similar to the Upgrade, the spacing between neighboring strings is approximately \SI{25}{\metre}, while the inter-DOM spacing is reduced, at \SI{1.5}{\metre}. These detector specifications are identical to those in~\cite{TheIceCube-Gen2:2016cap,PINGU}.

Independent of design details, both the Upgrade and PINGU will predominantly be sensitive to atmospheric neutrinos undergoing deep inelastic neutrino-nucleon scattering~\cite{Formaggio:2013kya}, at energies above a few \si{\giga\electronvolt}. The Cherenkov light that is emitted by charged final-state particles allows reconstructing the energy and zenith angle of the neutrino. Neutrino oscillation signatures from terrestrial matter effects, here especially the influence of Earth's core and lower mantle, have to be resolved in order to determine the NMO.

This paper builds on the original study in~\cite{BlennowSchwetz} which examines the NMO potential of a combined analysis of JUNO and PINGU, comparing it to the experiments' stand-alone capabilities. In order to test if the conclusions drawn in~\cite{BlennowSchwetz} hold under more detailed detector descriptions, we perform comprehensive livetime and parameter scans employing the same MC (where applicable) and the same set of systematic uncertainties that were used by the original design reports~\cite{JUNO,PINGU} of the two collaborations. We also perform a combined analysis that includes the Upgrade instead of PINGU and the 8- instead of the 10-core JUNO configuration.

Our paper is structured as follows: Chapter~\ref{chapter2} introduces the analysis framework and the modeling of JUNO, the Upgrade, and PINGU in our simulation. Chapter~\ref{chapter3} addresses the statistical approach to the NMO problem and discusses the benefits of performing a combined NMO analysis with JUNO and PINGU. It also demonstrates that similar conclusions hold in the case of a joint analysis of JUNO with 8 reactor cores and the Upgrade. We illustrate the origin of the synergy, contrast the projected combined sensitivities with the stand-alone performances of the respective experiments, and show how truth assumptions impact the NMO discovery potential. Chapter~\ref{chapter4} briefly examines the sensitivity to oscillation parameters of a combined analysis in case the NMO is correctly identified; the projected constraints are compared to the precision obtained by a recent global fit. Chapter~\ref{chapter5} summarizes our results and concludes with a short outlook.
\\
\section{Generation of expected spectra}
\label{chapter2}
We use the PISA~\cite{PISA} software to compute the expected experimental outputs of JUNO, the Upgrade, and PINGU. As depicted in Fig.~\ref{fig:PISAchain}, each experiment is modeled by means of a series of subsequent stages:

\begin{figure}[t]
\centering
\includegraphics[width=1.0\linewidth]{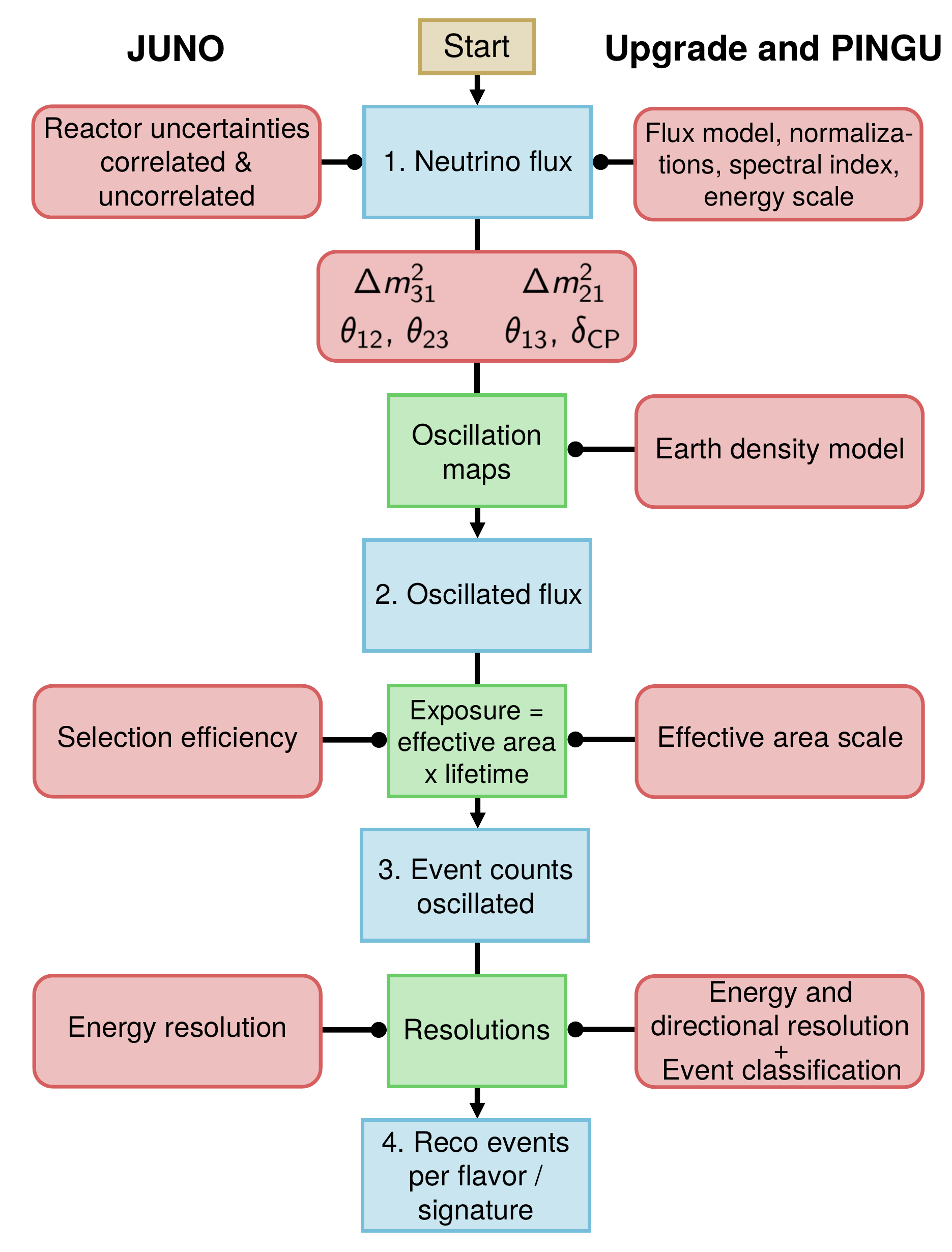}
\caption{\label{fig:PISAchain}Simulation chain employed in the modeling of the event distributions in JUNO and the IceCube Upgrade and PINGU. Blue boxes represent intermediate distributions, red ones physics inputs of the two experiments (JUNO on the left, the IceCube Upgrade and PINGU on the right, shared ones centered), whereas green boxes specify the entities that transform between the intermediate distributions. See text for details.}
\end{figure}

\begin{description}
\item[1. Flux] Outputs the energy- and direction-dependent neutrino flux at the location of the detector in the absence of oscillations.
\item[2. Oscillation] Calculates neutrino oscillation probabilities depending on energy and baseline given an input flux and weights the latter accordingly to obtain the oscillated neutrino flux at the detector.
\item[3. Exposure] Uses effective area/mass and livetime/operation time of the detector to convert the oscillated flux into an event number.
\item[4. Reconstruction and event classification] Smears the ``true'' event observables (energy and/or direction) with resolution kernels informed from detailed MC simulation (Upgrade and PINGU) or parametrized estimates (JUNO) and classifies Upgrade and PINGU events according to efficiencies obtained from separate, dedicated MC simulation.
\end{description}
We generate each experiment's event distribution as a histogram in 1d (JUNO) or 3d (Upgrade and PINGU). The common dimension is neutrino energy ($ E_{\nu} $). Upgrade and PINGU events are in addition binned in cosine-zenith ($ \cos(\theta_{Z}) $) and event type (``cascade-like'' or ``track-like'').

\subsection{JUNO}
Following the approach in~\cite{JUNO}, we use 200 bins equally spaced in energy covering the range $\SI{1.8}{\mega\electronvolt} < E_{\nu} < \SI{8.0}{\mega\electronvolt}$ for each reactor. For the flux and the oscillation stage we treat each reactor individually due to the different baselines ranging from \SI{52.12}{\kilo\metre} to \SI{52.84}{\kilo\metre}. Since JUNO does not distinguish neutrinos produced by different reactors, the binned energy spectra are superimposed after the exposure stage.

\paragraph{Flux}
The differential $\bar{\nu}_e$ flux per unit energy from one nuclear power plant's (NPP) reactor core at the detector location is approximated by the relation~\cite{JUNO}

\begin{equation} \label{equ:reactorflux}
\dv{\Phi_{\rm 0}}{E_\nu}\left(E_{\nu};L_\mathrm{core}\right)=\frac{W_{\rm th,core}}{L^2_\mathrm{core}\cdot\sum_{i}f_{i}e_{i}}\cdot\sum_{i}f_{i}\cdot S_{i}(E_{\nu})\text{ ,}
\end{equation}
where $ W_{\rm th,core} $ is the thermal power of the core, $L_\mathrm{core}$ its distance from the detector, and $ f_{i} $, $ e_{i} $, and $ S_{i}(E_{\nu}) $ are the fission fraction, the thermal energy released per fission, and the $\bar{\nu}_e$ energy distribution per fission for the $i$th isotope, respectively.

The values of $ f_{i} $ and $ e_{i} $ for the four dominant isotopes in an NPP reactor core are taken from~\cite{f}~and~\cite{e}. For the energy spectra $ S_{i}(E_{\nu}) $ we use the approximation given in~\cite{Isoflux}. In addition, each core's thermal power $W_{\rm th,core}$ and distance from the JUNO detector $L_\mathrm{core}$ are obtained from~\cite[Table~2]{JUNO}. This allows us to predict the unoscillated neutrino fluxes from all reactor cores at the location of the JUNO detector.

The systematic flux uncertainties we consider are an uncorrelated relative uncertainty of \SI{0.8}{\percent} on the thermal power of each core---proportional to the output neutrino flux---and a correlated uncertainty of \SI{2}{\percent} on all cores' overall neutrino flux normalization. These systematic error sources are implemented as Gaussian priors.

We do not include the so-called ``bump'' observed in various reactor anti-neutrino spectra~\cite{bump}: we expect the reference detector at the Taishan site, the recently proposed JUNO-TAO (Taishan Antineutrino Observatory)~\cite{JUNO_near,wonsak_bjoern_2018_1286850}, to provide a precise measurement of the unoscillated JUNO energy spectrum.

\paragraph{Oscillation}
The calculation of the $\bar{\nu}_e$ survival probability $\overline{P}_{ee}(E_\nu;L_\mathrm{core})$ for a given $\bar{\nu}_e$ energy $E_\nu$ and oscillation baseline $L_\mathrm{core}$ is performed using the Prob3++ code~\cite{prob3}. We include Earth matter effects, whose impact on the survival probability is at the \SI{1}{\percent} level for JUNO baselines~\cite{Li:2016txk}.

The oscillated $\bar{\nu}_e$ flux impinging on the JUNO detector from one reactor core at the distance $L_\mathrm{core}$ follows from the unoscillated flux $\Phi_0$ as

\begin{equation} \label{equ:reactorflux_osc}
\dv{\Phi_{\rm osc}}{E_\nu}\left(E_{\nu};L_{\rm core}\right)=\overline{P}_{ee}(E_{\nu}; L_{\rm core})\cdot\dv{\Phi_{\rm 0}}{E_\nu}\left(E_{\nu};L_\mathrm{core}\right)\text{ .}
\end{equation}

\paragraph{Exposure}
The rate of detected $\bar{\nu}_e$ events per unit energy is obtained as the sum over the contributions from all ten reactor cores,

\begin{equation} \label{equ:IBD spectrum}
\dv{\Dot{N}_{\rm IBD}}{E_{\nu}}\left(E_\nu\right)=\sum\limits_{\rm cores}\dv{\Phi_{\rm osc}}{E_\nu}\left(E_{\nu};L_{\rm core}\right)\cdot A_{\rm eff}(E_\nu) \text{ ,}
\end{equation}
where we have dropped the explicit dependence on the set of baselines $\{L_\mathrm{core}\}$ on the left. $A_{\mathrm{eff}}(E_\nu)$ is the detector's \emph{effective area}, which corresponds to the product of the number of target protons $ N_{p} $ and the IBD cross section $ \sigma_{\rm IBD}(E_\nu) $, corrected by the selection efficiency $\epsilon$,
\begin{equation}\label{equ: juno aeff}
A_{\mathrm{eff}}(E_\nu)= N_{p}\cdot\sigma_{\rm IBD}(E_{\nu})\cdot \epsilon\text{ .}
\end{equation}
We evaluate Eq.~(\ref{equ: juno aeff}) assuming $ N_{p}=\SI{1.54e33}{} $ (chosen by matching the IBD event rate to that reported in~\cite{JUNO}, see below) and substituting the IBD cross section $\sigma_\mathrm{IBD}(E_\nu)$ from~\cite{Vogel:1999zy}. Eq.~(\ref{equ:IBD spectrum}) gives the differential event spectrum \emph{before} reconstruction. Without selection cuts ($\epsilon=1$), integrating Eq.~(\ref{equ:IBD spectrum}) over the considered energy range yields $\Dot{N}^{\epsilon=1}_\mathrm{IBD}\sim \SI{83}{\day^{-1}}$~\cite{JUNO} for the 10-core JUNO configuration. Including selection cuts, the IBD selection efficiency $\epsilon$ is given as $\SI{73\pm1}{\percent}$~\cite[Table~3]{JUNO}. This results in a $\bar{\nu}_e$ rate of $\Dot{N}_\mathrm{IBD}\sim \SI{60}{\day^{-1}}$. In order to account for reactor and detector downtime, we assume an effective livetime of 300 days per year of data taking~\cite{Li:2013zyd}.

\paragraph{Reconstruction}
JUNO determines the neutrino energy via the visible energy $E_\mathrm{vis}$ of the prompt IBD positron, which corresponds to the neutrino energy reduced by \SI{0.784}{\mega\electronvolt}: $E_\mathrm{vis} = E_\nu - \SI{0.784}{\mega\electronvolt}$. Since JUNO aims to achieve an IBD positron visible-energy resolution of $ \SI{3}{\percent} $ at $ E_\mathrm{vis} = \SI{1}{\mega\electronvolt} $, each spectral bin is ``smeared'' with the corresponding energy uncertainty

\begin{equation}
\Delta E(E_\nu)/\si{\mega\electronvolt}=\SI{3}{\percent}\cdot\sqrt{E_{\nu}/\si{\mega\electronvolt}-0.784}
\label{eq:eres}
\end{equation}
by convolving the true event distribution in $E_\nu$ with a normal distribution of standard deviation $ \Delta E(E_\nu) $.

\paragraph{Background sources}
Several background sources contribute to JUNO's event distribution. Besides the two long-baseline nuclear power plants Daya Bay and Huizhou, accidental backgrounds, fast neutrons, $ ^{9}$Li/$^{8}$He decays, $ ^{13}$C$(\alpha,n)^{16}$O reactions, and geoneutrinos are relevant. All of these can mimic a reactor $\bar{\nu}_e$ IBD event.\\
Daya Bay and Huizhou~\cite[Table~2]{JUNO} are added to the simulation in the same manner as the signal-producing reactor cores, leading to 5 additional selected reactor $\bar{\nu}_e$ events per day in JUNO. The rates of the other backgrounds are extracted from~\cite[Fig.~19]{JUNO} and are added after reconstruction. Note that the $ ^{9}$Li/$^{8}$He rate is reduced by \SI{97.7}{\percent}~\cite{JUNO} due to the application of a cosmogenic veto. The corresponding loss of IBD events is included in the selection efficiency quoted above.

\paragraph{Systematics summary}
Table~\ref{tab:JUNOsys} summarizes all the systematic parameters applied in the simulation of the JUNO event distribution. Here, the uncorrelated shape uncertainty of the output event histogram is realized through a modification of the $ \chi^{2} $ definition which we minimize numerically in the NMO analysis (see Eq.~(\ref{eq:JUNOchi2})).

\begin{table}[H]
  \centering
    \begin{tabular}{lcc}
    \toprule
    \textbf{systematic error source} & \textbf{\boldmath{uncertainty}} & \textbf{fit}\\
    \midrule
    uncorrelated reactor flux normalization & $\SI{0.8}{\percent}$ & \checkmark\\
    correlated reactor flux normalization & $\SI{2}{\percent}$ & \checkmark\\
    IBD selection efficiency & $ \SI{1}{\percent} $ & \checkmark\\
    uncorrelated shape uncertainty & $ \SI{1}{\percent} $ & $\times$\\
    \bottomrule
    \end{tabular}%
  \caption{Systematic error sources employed in the modeling of the JUNO experiment, together with their assumed $1\sigma$ Gaussian uncertainties. The third column specifies whether each systematic error is implemented as a free fit parameter (checkmark) or not (cross).}
  \label{tab:JUNOsys}
\end{table}%

\begin{figure}
\centering
\input{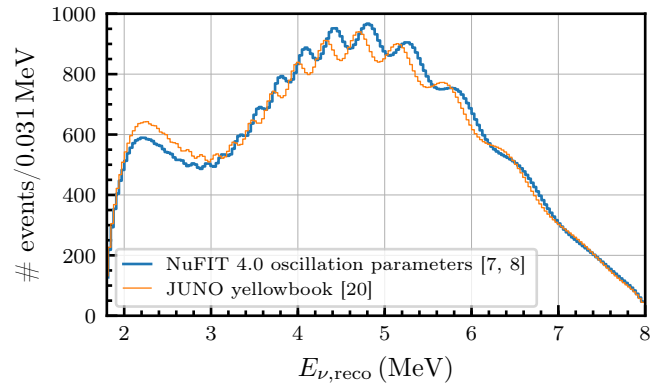}
\caption{\label{fig:juno_out} Expected distribution of reconstructed energies $E_{\nu,\mathrm{reco}}$ of reactor $\bar{\nu}_e$ events in JUNO given true NO for our analysis binning after 6 years of operation (1800 days), without any background contribution. Shown is the expected spectrum assuming the nominal oscillation parameter inputs from~\cite{JUNO} (thin orange line) and the spectrum assuming the inputs from Table~\ref{tab:usedParams} (thick blue line).}
\end{figure}

In a stand-alone analysis which assumes the same oscillation parameter inputs as~\cite{JUNO}, we obtain an NMO significance of $3.2\sigma$ for true NO after 6 years of livetime, compared to $3.3\sigma$ in~\cite[Table~4 (``standard sens.'')]{JUNO}. Due to more recent global inputs for the oscillation parameters~\cite{Nufit4.0,nufit4.0url} (reproduced in Table~\ref{tab:usedParams} for convenience), the analysis presented in this work finds a nominal stand-alone JUNO sensitivity that is reduced\footnote{The main contributions to this reduction are brought about by shifts in $\Delta m_{31}^{2}$ and $\Delta m_{21}^{2}$.} to $2.8\sigma$ after 6 years. Assuming the NO to be true, Fig.~\ref{fig:juno_out} compares the expected $\bar{\nu}_e$ energy spectrum (without any backgrounds) after 6 years resulting from these recent global inputs to the one based on the oscillation parameter assumptions made in~\cite{JUNO}.

\paragraph{Reduced reactor $\bar\nu_e$ flux} Since it is possible that two of the ten nominal JUNO signal reactor cores (Taishan 3 \& 4) will not be in operation at the start of data taking, we also study a conservative scenario in which only the remaining eight are available. In this case, the signal reactor $\bar\nu_{e}$ rate after selection is reduced to about \SI{74}{\percent} of its nominal value ($\sim \SI{60}{\day^{-1}}\rightarrow\;\sim \SI{44}{\day^{-1}}$). Given that this setup does not represent the nominal assumptions regarding the JUNO source configuration, there have been no corresponding sensitivity studies prior to this work.

\subsection{The IceCube Upgrade and PINGU}
In the case of the Upgrade we choose 10 bins linear in the reconstructed neutrino cosine zenith $ \cos(\theta_{Z,\mathrm{reco}}) $, covering the range $ -1 \leq \cos(\theta_{Z,\mathrm{reco}}) \leq 0 $ of Earth-crossing (from vertically up-going to horizontal) trajectories. Reconstructed neutrino energies $E_{\nu,\mathrm{reco}}$ are binned using 10 logarithmically spaced bins spanning $ \SI{3}{\giga\electronvolt} \leq E_{\nu,\mathrm{reco}} \leq \SI{80}{\giga\electronvolt} $. Due to the better resolutions of PINGU compared to the Upgrade, the number of bins in energy is raised to 39, with the lower bound reduced to $ E_{\nu,\mathrm{reco}}= \SI{1}{\giga\electronvolt}$, in analogy to~\cite{PINGU}. Similarly, the number of bins in cosine zenith is raised to 20.

\paragraph{Flux}
Atmospheric neutrino fluxes at the detector site at the South Pole are obtained from tabulated MC simulations performed by Honda et al.~\cite{Honda}. For each neutrino type,\footnote{In this section, ``neutrino'' refers to both ``neutrino'' and ``anti-neutrino'' unless the latter is used explicitly.} these tables provide average unoscillated fluxes across bins in neutrino energy, cosine zenith angle, and azimuth. Starting from two-dimensional---azimuth- and season-averaged---tables, we evaluate the flux of each atmospheric neutrino species on a $(200\times390)$ grid in $(\cos(\theta_Z), E_\nu)$ via an integral-preserving interpolation~\cite{PISA}. Systematic flux uncertainties which we consider include the neutrino energy scale, to which we assign an uncertainty of \SI{10}{\percent}, the ratio between electron and muon neutrino fluxes, assuming an uncertainty of \SI{3}{\percent}, the ratio between neutrino and anti-neutrino fluxes, with an uncertainty of \SI{10}{\percent}, as well as a shift of the spectral index with respect to the nominal flux model, assuming an uncertainty of \SI{5}{\percent}.

\paragraph{Oscillation} \label{matter-osc}
The NMO signature in IceCube depends on the Earth matter density distribution through its impact on the neutrino oscillation probabilities~\cite{Wolfenstein:1978ue,Mikheev:1986wj,Petcov:1986qg,Akhmedov:2006hb}. In order to calculate the latter we again use the Prob3++ code together with the Preliminary Reference Earth Model (PREM)~\cite{prem}, with a division of the Earth into 12 constant-density layers. The electron fractions in the Earth's core and mantle are assumed to be $ Y_{e}^{\rm c}=0.4656 $ and $ Y_{e}^{\rm m}=0.4957 $, respectively. The oscillation baseline for a given trajectory is determined by assuming a neutrino production height in the atmosphere of \SI{20}{\kilo\metre}~\cite{PISA} and a detector depth of \SI{2}{\kilo\metre}. Together with the Earth's radius of about \SI{6371}{\kilo\metre} and the neutrino zenith angle, these fully determine the neutrino trajectory and thereby the traversed density profile, which in turn is required to calculate oscillation probabilities for a given neutrino energy. Since the intrinsic atmospheric flux of tau neutrinos is negligible at the energies of interest to the NMO~\cite{Lee:2004zm}, the relevant transition channels that need to be considered are $\nu_e \to \nu_{e,\mu,\tau}$ and $\nu_\mu \to \nu_{e,\mu,\tau}$. The oscillated flux (in a bin centered on $\log_{10} E_\nu/\si{\giga\electronvolt}$ and $\cos(\theta_Z)$) of atmospheric neutrinos of flavor $\beta$ impinging on the detector is then obtained from the sum over the unoscillated fluxes $\Phi_{0,\alpha}(E_\nu,\cos(\theta_Z))$ of all initial flavors $\alpha$, weighted by the probabilities $P_{\alpha\beta}(E_\nu,\cos(\theta_Z))$ of the flavor transitions $\alpha\to\beta$ (note that we drop the explicit dependence on energy and cosine zenith for brevity on both sides of Eqns.~(\ref{eq:osc atmospheric flux}),~(\ref{eq:atmospheric event rate})):
\begin{equation}
    \Phi_{\mathrm{osc},\beta} = \sum_\alpha P_{\alpha\beta} \cdot \Phi_{0,\alpha}\text{ .}
    \label{eq:osc atmospheric flux}
\end{equation}

\paragraph{Exposure}
Given an incoming oscillated flux of neutrinos of flavor $\beta$, $\Phi_{\mathrm{osc},\beta}$, the resulting event rate for a given interaction channel (charged current (CC) or neutral current (NC)),

\begin{equation}
    \Dot{N}_{\alpha} =\Phi_{\mathrm{osc},\beta}\cdot A_{\mathrm{eff},\beta}\cdot s_A\text{ ,}
    \label{eq:atmospheric event rate}
\end{equation}
is calculated for each bin in true neutrino energy $E_\nu$ and cosine zenith $\cos(\theta_Z)$. Here, $A_{\mathrm{eff},\beta}\left(E_\nu,\cos(\theta_{Z})\right)$ is the effective area of the detector for the given neutrino species and interaction type. The scaling factor $s_A$ represents a universal systematic uncertainty on all effective areas.

The effective area functions $A_{\mathrm{eff},\beta}$ for both the Upgrade and PINGU are obtained by binning MC events from detailed detector simulations~\cite{PINGU} weighted by their individual effective-area weights on a $(20\times39)$ grid in $(\cos(\theta_Z), E_\nu)$. In order to mitigate fluctuations due to limited MC statistics, the histograms are smoothed in both dimensions using Gaussian smoothing followed by spline smoothing~\cite{PISA}. We extract separate effective areas for neutrinos and anti-neutrinos. We similarly separate the individual flavors in the case of CC interactions. Events due to NC interactions, however, are combined into a single effective-area function. They are also treated identically in reconstruction and event classification (see below). Hence, we are left with the event categories $ \nu_{e} $ CC, $ \nu_{\mu} $ CC, $ \nu_{\tau} $ CC, $ \nu_{e,\mu,\tau} $ NC (and the same for anti-neutrinos). The effective area scale $s_A$ serves as a normalization of the combined neutrino and anti-neutrino event rate. It is assigned an uncertainty of \SI{10}{\percent}. Based on an uptime of the IceCube detector that routinely exceeds \SI{99}{\percent}~\cite{Aartsen:2016nxy}, the assumed effective livetime for the Upgrade and PINGU is 365.25 days per year of data taking.

\paragraph{Reconstruction and event classification}
Having undergone a detailed likelihood reconstruction at the single photon level, the same sets of MC events used to model the Upgrade's and PINGU's effective areas are employed in applying reconstruction resolutions. Here, variable bandwidth kernel density estimation (VBWKDE) mitigates the effects of statistical fluctuations and yields the probability distributions (``smearing kernels'') that map from true neutrino energy and cosine zenith onto the corresponding reconstructed observables~\cite{PISA}. The third output dimension is the event class, ``cascade-like'' or ``track-like''. It is determined from a pre-calculated score with discrimination power between $\nu_\mu + \bar{\nu}_\mu$ CC events, which contain a minimum-ionizing muon track, and the rest, which result in a rather spherical light deposition pattern~\cite{PINGU}.
Neutrinos and anti-neutrinos are grouped together before the resolution and classification distributions are generated.

\paragraph{Background sources}
Since the Earth acts as a shield against up-going cosmic ray muons, and strict cuts are applied to reject the muon flux entering through the surrounding IceCube detector from above, we do not include any background contamination in the event distributions of either the Upgrade or PINGU~\cite{PINGU}.

\paragraph{Systematics summary}
Table~\ref{tab:PINGUsys} provides an overview of the systematic uncertainties we consider for the Upgrade and PINGU. Both the energy and effective area scale have been introduced to capture the effects of multiple systematic uncertainties. This is a consequence of the significant amount of MC simulation that would otherwise be required for a detailed study of the various detector-related systematics, such as the overall detection efficiency of the optical modules~\cite{PINGU}. Furthermore, data taken with the Upgrade or PINGU will profit from an improved knowledge of the optical properties of the ice, owing to the deployment of novel in-situ calibration devices and a modified hole-drilling method~\cite{upgrade_icrc:2019}. As a result, we do not expect a degradation of the NMO sensitivity due to ice uncertainties. For a discussion of their impact on the PINGU NMO sensitivity and the NMO measurement with IceCube DeepCore, see also \cite{PINGU} and \cite{Aartsen:2019eht}, respectively.

\begin{table}[t]
  \centering
    \begin{tabular}{lcc}
    \toprule
    \textbf{systematic error source} & \textbf{\boldmath{uncertainty}} & \textbf{fit}\\
    \midrule
    atmospheric spectral index shift & $\SI{5}{\percent}$ & \checkmark\\
    $ \nu_{e}/\nu_{\mu} $ flux ratio scale & $\SI{3}{\percent}$ & \checkmark \\
    $ \nu/\bar{\nu} $ flux ratio scale & $\SI{10}{\percent}$ & \checkmark\\
    energy scale & $\SI{10}{\percent}$ & \checkmark\\
    effective area scale & $\SI{10}{\percent}$ & \checkmark\\
    \bottomrule
    \end{tabular}%
  \caption{Systematic error sources employed in the modeling of the Upgrade and PINGU. The entries are to be interpreted in the same way as those of Table~\ref{tab:JUNOsys}.}
  \label{tab:PINGUsys}%
\end{table}%

With the oscillation parameter values assumed in~\cite{PINGU}, our stand-alone analysis of PINGU yields a median significance\footnote{For the sake of comparison, in this paragraph only we employ the definition of the median NMO sensitivity given in~\cite[Eq.~(A.13)]{PINGU}, which differs slightly from the ``standard'' sensitivity proxy adopted throughout the remainder of this paper. Note that the sensitivities in this paragraph should not be compared to those shown later in Table~\ref{tab:sens} due to the deviating underlying assumptions.} of $3.2\sigma$ to exclude the IO for true NO after 4 years of data taking, somewhat larger than the value of $2.8\sigma$ presented in~\cite{PINGU}, mostly due to the improved VBWKDE treatment of the resolution functions~\cite{PISA}. For the Upgrade, in contrast, the are no previously published sensitivities.

\begin{figure}[t]
\centering
\input{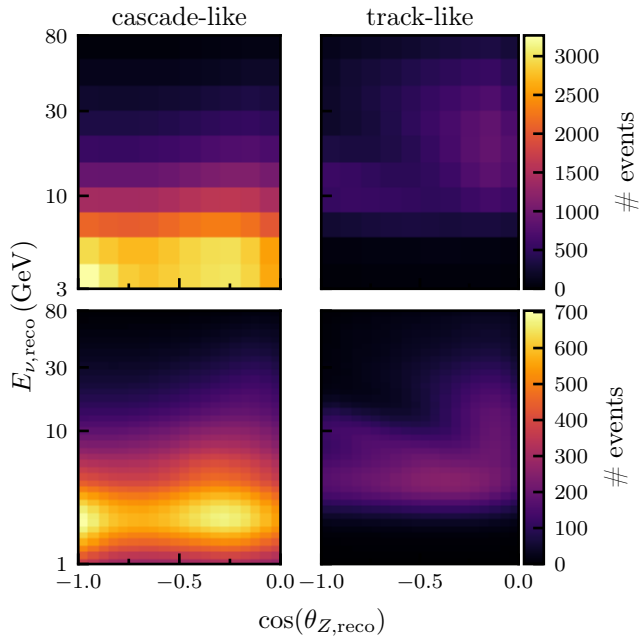}
\caption{\label{fig:pingu_out} Nominal expected event distributions given true NO for our analysis binning in reconstructed neutrino cosine zenith $\cos(\theta_{Z,\mathrm{reco}})$ and reconstructed neutrino energy $E_{\nu,\mathrm{reco}}$ for the IceCube Upgrade (top) and PINGU (bottom) after 4 years of operation.}
\end{figure}

Figure~\ref{fig:pingu_out} shows the expected event distribution for 4 years of data taking given the more recent global NO inputs from Table~\ref{tab:usedParams} for the Upgrade as well as for PINGU.

\section{Expected NMO Sensitivities in a combined analysis}
\label{chapter3}

\subsection{Statistical approach}
\label{subsec:statistics}
\paragraph{Sensitivity proxy}
In the following, all NMO sensitivities are obtained by producing toy data under the ordering assumed to be true (TO)---generated from the oscillation parameter values in Table~\ref{tab:usedParams} (unless stated otherwise) and with all systematic uncertainties at their nominal values---and fitting the toy data by numerically minimizing a $\chi^2$ function over all free parameters while restricting the minimizer to the wrong $\Delta m^2_{31}$ half-plane (wrong ordering, WO). Since no statistical fluctuations are applied to the toy data, we refer to it as the ``Asimov dataset''~\cite{asimov}. We employ the test statistic
\begin{equation}
    \overline{\Delta \chi^2} \equiv |\chi^2_\mathrm{NO} - \chi^2_\mathrm{IO}|\text{ ,}
\end{equation}
where $\chi^2_{\mathrm{NO}(\mathrm{IO})} \equiv \mathrm{min}_{\{p_i\} \in \mathrm{NO}(\mathrm{IO})} \chi^2$ corresponds to the minimum when the fit hypothesis (with free parameters $\{p_i\}$) is NO (IO). In the Asimov approach, toy data generated under the TO results in $\chi^2_\mathrm{TO}=0$ by construction. Hence, one has
\begin{equation}
    \overline{\Delta \chi^2} = \begin{cases}
\chi^2_\mathrm{IO} & \text{true NO} \\
\chi^2_\mathrm{NO} & \text{true IO}
\end{cases}\text{ .}\label{eq:Asimov delta chisquare}
\end{equation}
We then convert Eq.~(\ref{eq:Asimov delta chisquare}) into a median NMO significance by taking $\sqrt{\overline{\Delta \chi^2}}$. This sensitivity proxy corresponds to a one-sided number of standard deviations at which the WO is excluded. Note that the relation is exact in the case of symmetric Gaussian distributions of the $\Delta \chi^2$ test statistic~\cite{n_sig}, which we have found to hold to a good approximation.\footnote{See also~\cite{JUNO,PINGU,n_sig} for the test statistic distributions observed in PINGU and JUNO.}

\begin{table}[t]
   \centering
     \begin{tabular}{cccccc}
     \toprule
     \textbf{parameter} & \textbf{true NO} & \textbf{true IO} & \textbf{fit} & \textbf{fit range}\\
     \midrule
     $ \Delta m_{31}^{2}\;(\si{\electronvolt\squared}) $ & \SI{2.53e-3}{} & \SI{-2.44e-3}{} & \checkmark & nom. $\pm 3\sigma$ \cite{Nufit4.0,nufit4.0url}\\
     $ \Delta m_{21}^{2}\;(\si{\electronvolt\squared})  $ & \multicolumn{2}{c}{\SI{7.39e-5}{}} & $\times$ & - \\
     $ \theta_{12}\;(\si{\degree}) $ & \multicolumn{2}{c}{33.82} & \checkmark & nom. $\pm 3\sigma$ \cite{Nufit4.0,nufit4.0url}\\
     $ \theta_{13}\;(\si{\degree}) $ & $8.61\pm0.13$ & $8.65\pm0.13$ & \checkmark & nom. $\pm 3\sigma$ \cite{Nufit4.0,nufit4.0url}\\
     $ \theta_{23}\;(\si{\degree}) $ & $ 49.6 $ & $ 49.8 $ & \checkmark & nom. $\pm  3\sigma$ \cite{Nufit4.0,nufit4.0url}\\
     $ \delta_{\rm CP}\;(\si{\radian})$ & 0 & 0 & $\times$ & -\\
     \bottomrule
     \end{tabular}%
   \caption{Nominal input oscillation parameter values employed in our NMO analysis. The central values for all but $\delta_{\rm CP}$ are taken from a fit to global data available at the end of 2018~\cite{Nufit4.0,nufit4.0url}. The second-to-last column (``fit'') denotes whether the given parameter is fit (checkmark) or kept fixed (cross). In case it is fit, the last column shows the range explored by the minimizer with respect to the nominal parameter value.}
   \label{tab:usedParams}%
\end{table}%

\paragraph{Nomenclature}
In the following, \emph{combined fit} refers to the minimization of the expression

\begin{equation}
\chi^{2} \equiv \chi^{2}_{\mathrm{JUNO,stat}}+\chi^{2}_{\mathrm{IceCube,stat}}+\chi^2_{\mathrm{prior}}\text{ ,}
\end{equation}
\noindent where
\begin{align}
\chi^{2}_{\mathrm{JUNO,stat}} &=
  \sum_{i=1}^{N^{\text{JUNO}}_{\mathrm{bins}}}\frac{(n^{\mathrm{obs}}_{i}-n^{\mathrm{exp}}_{i})^{2}}{(\sigma_\mathrm{uncorr}\cdot n^{\mathrm{exp}}_{i})^{2}+n^{\mathrm{exp}}_{i}}\text{ ,}\label{eq:JUNOchi2}\\
\chi^{2}_{\mathrm{IceCube,stat}} &= \sum_{j=1}^{N^{\mathrm{IceCube}}_{\mathrm{bins}}}\frac{(n^{\mathrm{obs}}_{j}-n^{\mathrm{exp}}_{j})^{2}}{n^{\mathrm{exp}}_{j}}\text{ ,}\label{eq:PINGUchi2}\\
\chi^2_{\mathrm{prior}} &= \sum_{k=1}^{N_\mathrm{prior}} \frac{\left(\Delta p_k\right)^2}{\sigma^2_k}
\label{eq:priorchi2}\text{ .}
\end{align}

In the three $\chi^2$ contributions~(\ref{eq:JUNOchi2})--(\ref{eq:priorchi2}), the indices $i$ in Eq.~(\ref{eq:JUNOchi2}) and $j$ in Eq.~(\ref{eq:PINGUchi2}) run over the event histogram bins employed for JUNO and IceCube (either Upgrade or PINGU), respectively, whereas $k$ in Eq.~(\ref{eq:priorchi2}) runs over all nuisance parameters subject to external Gaussian constraints. $n^{\rm obs}$ is the measured number of events in the toy data, $n^{\rm exp}$ the expected number, and $\sigma_\mathrm{uncorr}=0.01$ the uncorrelated reactor flux shape uncertainty. For the $k$th nuisance parameter subject to an external constraint, a deviation $\Delta p_k$ from its nominal value is penalized according to the parameter's standard deviation $\sigma_k$ as $\left(\Delta p_k\right)^2/\sigma^2_k$. 

The set of nuisance parameters considered depends on the experimental configuration. In the case of JUNO, these are the oscillation parameters $\Delta m_{31}^{2}$, $\theta_{12}$, $\theta_{13}$, and the systematic uncertainties enumerated in Table~\ref{tab:JUNOsys}. In the case of IceCube, they are $\Delta m_{31}^{2}$, $\theta_{13}$, $\theta_{23}$, and the systematic uncertainties in Table~\ref{tab:PINGUsys}. A combined fit includes the union of these two sets of parameters. Oscillation parameters that are kept fixed in the stand-alone or combined fits have little to no impact on the NMO sensitivities we find.

In some cases, the significance derived from evaluating Eq.~(\ref{eq:JUNOchi2})~or~(\ref{eq:PINGUchi2}) at the minimum of the combined fit is given in addition; we refer to it as a given experiment's \emph{statistical contribution} to the combined significance.

Finally, whenever we show fit parameter scans within the TO, the definition of $\overline{\Delta \chi^2}$ differs from that given by Eq.~(\ref{eq:Asimov delta chisquare}) in that $\overline{\Delta \chi^2}$ corresponds to the local $\chi^2_\mathrm{TO}$ value (at the considered point within the TO hypothesis parameter space), as opposed to the local $\chi^2_\mathrm{WO}$ value in the case of fit parameter scans within the WO.

\subsection{Synergy effects}
\label{sec:synergy}
\begin{figure*}
\centering
\input{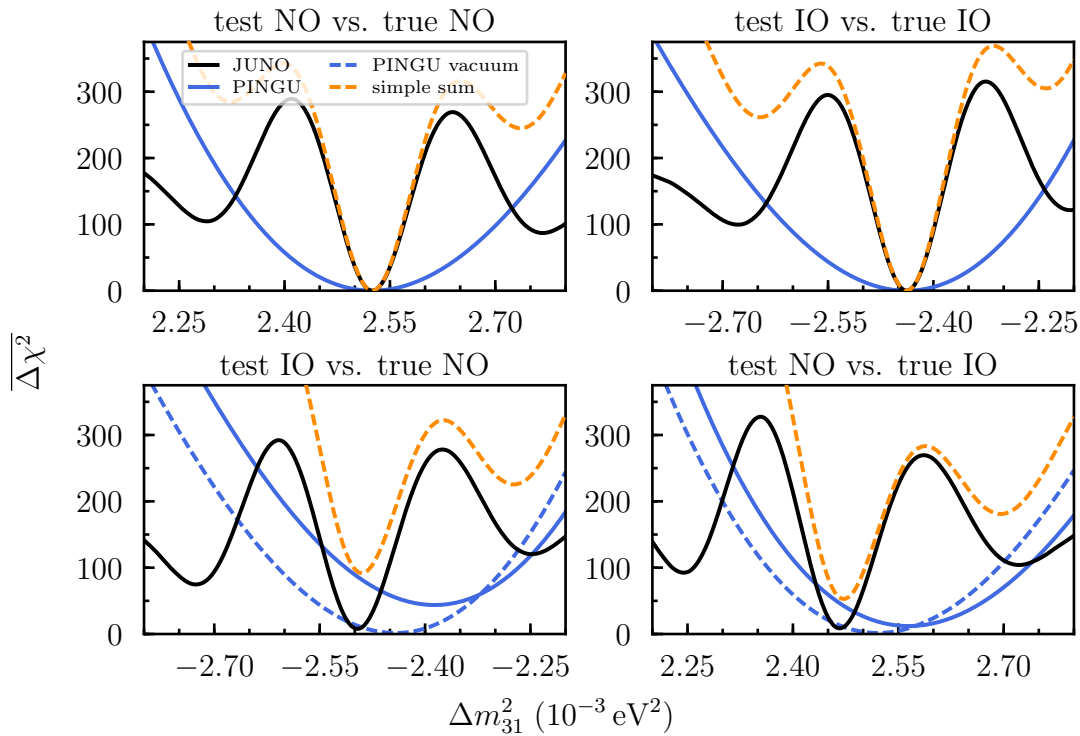}
\caption{$\overline{\Delta\chi^2}$ profiles as function of the tested/fit values of $ \Delta m_{31}^{2} $ within the true ordering (upper panels) and the wrong ordering (lower panels) for a livetime of 6 years for both experiments. On the left (right), the NO (IO) is assumed to be true. The scans within the wrong ordering illustrate the synergy effect of performing a combined fit. Here, a tension in the best-fit values of $ \Delta m_{31}^{2} $ for PINGU and JUNO is visible that is greater than the ``resolution'' of the two experiments. Shown in addition are the hypothetical wrong-ordering profiles assuming a vanishing matter density along all neutrino trajectories in the case of PINGU (labeled ``PINGU vacuum''). The line labeled ``simple sum'' (dashed orange) is the sum of the ``JUNO'' and ``PINGU'' curves at each tested value of $\Delta m^2_{31}$.}
\label{fig:synergy}
\end{figure*}

\begin{figure*}[t]
\centering
\input{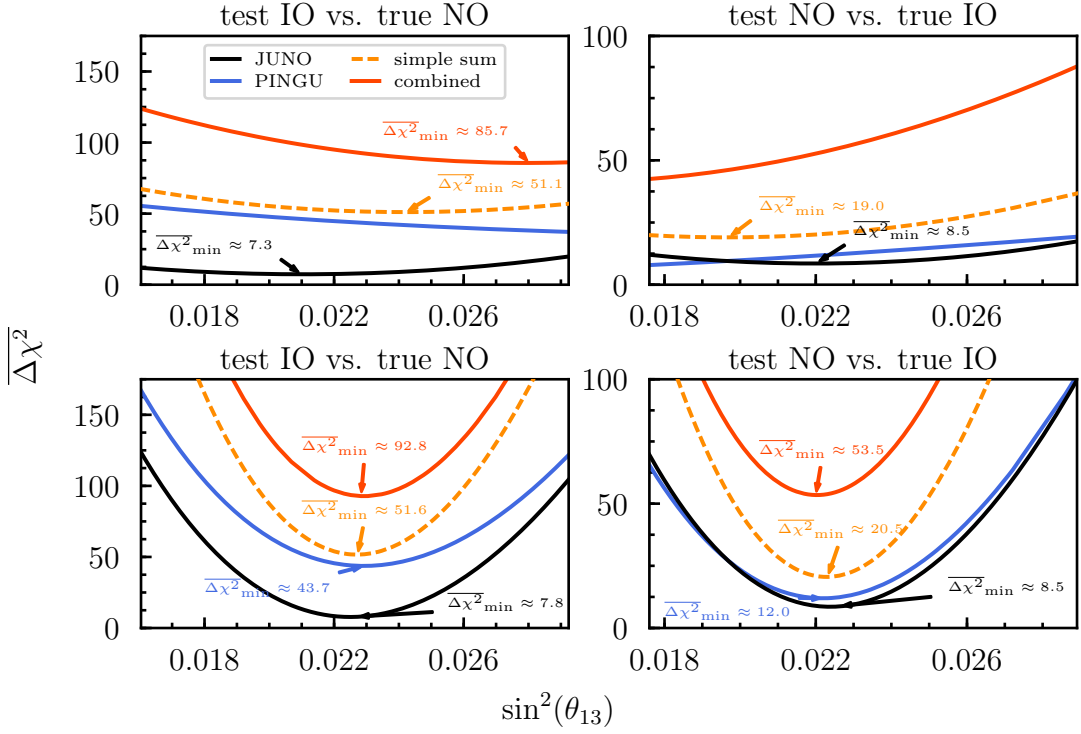}
\caption[$ \theta_{13} $ synergy effect]{$ \theta_{13} $ synergy effect after 6 years of livetime of both experiments, without (top) and with (bottom) a prior on the parameter, assuming true NO on the left and true IO on the right. In each case, the wrong ordering is fit to the true one. In contrast to Fig.~\ref{fig:synergy}, here we show the full combined fit (solid red) in addition to the ``simple sum'' (dashed orange).}
\label{fig:theta_synergy}
\end{figure*}
Why is it to be expected that the NMO sensitivity of the combined analysis exceeds the simple sum of the two individual experiments' sensitivities? Here, we discuss PINGU in the combined analysis with JUNO with the aim of clarity in illustrating the synergy effects and to be able to compare our results to~\cite{BlennowSchwetz}.

The upper part of Fig.~\ref{fig:synergy} shows the JUNO and PINGU $\overline{\Delta\chi^2}$ profiles as function of the tested value of $ \Delta m_{31}^{2} $ within the \emph{true} ordering (NO on the left, IO on the right) for an exposure time of 6 years, minimized over all other parameters at each point. As the scan is performed on the Asimov dataset, the minimal $ \overline{\Delta\chi^2} $ value is found at the true value of $ \Delta m_{31}^{2} $ and vanishes exactly for both JUNO and PINGU; the widths of the minima give the precision to which the two experiments are able to constrain $ \Delta m_{31}^{2} $ when either the NO or the IO are correctly identified.

The result changes profoundly when the same scan is performed while assuming the \emph{wrong} ordering in the fit, depicted in the lower panels of Fig.~\ref{fig:synergy}. The inherent tension between data and hypothesis, i.e., the wrong sign of $\Delta m^2_{31}$, in both JUNO and PINGU results in $ \overline{\Delta\chi^2} > 0$ for all tested values of $ \Delta m_{31}^{2} $. In this case, each experiment's minimum $\overline{\Delta\chi^2}$ serves as a sensitivity proxy---the larger it is, the better each experiment distinguishes the wrong ordering from the true one.

Most importantly, the $ \Delta m_{31}^{2} $ values at the minima (the {\it best-fit} values) no longer coincide. For both experiments, in terms of the absolute value $|\Delta m_{31}^2|$ the best fit is smaller than the truth when NO is true and greater when IO is true. While this deviation is always larger for PINGU (and increased by Earth matter effects compared to the hypothetical case of vacuum oscillations), the JUNO fit provides a more pronounced $ \overline{\Delta\chi^2} $ minimum, which in addition is much narrower than the shift between the two experiments' best fits.
It is precisely this configuration which explains the benefits of the combined analysis: by forcing the simultaneous fit of JUNO's and PINGU's event distributions to assume the same value for $\Delta m^2_{31}$ within the wrong NMO, the $\overline{\Delta\chi^2}$ minimum grows much beyond the simple sum of the two individual fits' minima. Note that in the absence of tension in the non-displayed nuisance parameters (which is approximately the case here), the best fit in $\Delta m^2_{31}$ would simply be given by the position of the minimum of the sum of the two experiments' $ \overline{\Delta\chi^2} $ profiles (labeled ``simple sum'' in Fig.~\ref{fig:synergy}); the combined NMO sensitivity would be given by the value of $\overline{\Delta \chi^2}$ at the minimum.

The differing $ \Delta m_{31}^{2} $ dependencies of the NMO measurements by JUNO and PINGU constitute the most pronounced synergy effect of their combined analysis. Similar to the lower part of Fig.~\ref{fig:synergy}, Fig.~\ref{fig:theta_synergy} shows the $\overline{\Delta\chi^2}$ profiles as a function of the tested value of $ \sin^2(\theta_{13}) $ in the wrong ordering, that is, when only allowing the minimizer to explore the region $\Delta m^2_{31} < 0\,(> 0)$ for true NO (IO) at each value of $\sin^2(\theta_{13})$. Here, we only subject $\theta_{13}$ to the prior of Table~\ref{tab:usedParams} in the lower panels; no prior is used in the upper panels.

In absence of a prior, the sum of the minimal $\overline{\Delta\chi^2}$ of the stand-alone profiles is significantly lower than the minimum of the summed $\overline{\Delta\chi^2}$ curve. The latter in this case represents a combined analysis that only forces the two experiments to assume the same $ \theta_{13} $ value within the wrong NMO. Both the NMO sensitivity derived from summing the minima and from the joint fit of $\theta_{13}$ only are significantly smaller than that obtained in a full joint analysis (``combined''), in which also the values of all other oscillation parameters are required to match. This in turn clearly indicates that the tension in $\theta_{13}$ is small compared to that in $\Delta m_{31}^2$.

Even the comparably small beneficial impact due to $\theta_{13}$ mostly disappears once the latter is assigned the prior from Table~\ref{tab:usedParams}, as shown in the lower panel of Fig.~\ref{fig:theta_synergy}. Here, it becomes apparent that summing the stand-alone minima yields approximately the same NMO sensitivity as the joint fit of $\theta_{13}$ only.

We consider the prior important for studying realistic scenarios and apply it in all other sensitivity calculations for two reasons. PINGU's best-fit values for $ \theta_{13} $ within the wrong ordering are far outside the globally allowed regions~\cite{Nufit4.0,nufit4.0url} (and outside the plotting ranges of the top panels of Fig.~\ref{fig:theta_synergy}). Also, its $\overline{\Delta\chi^2}$ minima in absence of the prior are visibly reduced.

The fact that IceCube is not sensitive to $ \theta_{12} $ and JUNO is not sensitive to $ \theta_{23} $ means that these parameters do not contribute any additional synergies. Moreover, we expect no other synergy effects to benefit the combined analysis since the two experiments are not assumed to share any systematic uncertainties beside the oscillation parameters discussed above.

\subsection{Sensitivity after 6 years of exposure}
\label{sec:6years}

\setlength{\tabcolsep}{1em}
\begin{table*}
  \centering
    \begin{tabular}{l||cc|cc}
    \toprule
    \textbf{} & \textbf{JUNO (eight cores)} & \textbf{Upgrade} & \textbf{JUNO} & \textbf{PINGU}\\
    \midrule
    stand-alone & $2.4\sigma/2.5\sigma$ & $3.8\sigma/1.8\sigma$  & $2.8\sigma/2.9\sigma$ & $6.6\sigma/3.5\sigma$\\
    statistical contribution & $2.6\sigma/2.6\sigma$ & $5.8\sigma/4.3\sigma$ & $3.2\sigma/3.3\sigma$ & $8.5\sigma/6.3\sigma$ \\
    combined analysis & \multicolumn{2}{c|}{$6.5\sigma/5.1\sigma$} & \multicolumn{2}{c}{$9.6\sigma/7.3\sigma$} \\
    \bottomrule
    \end{tabular}%
  \caption{Expected NMO sensitivities (true NO/true IO) after 6 years of operation, for the stand-alone experiments as well as for their combined analysis (JUNO with eight cores and the IceCube Upgrade, JUNO and PINGU), including the statistical contributions within the latter.}
  \label{tab:sens}%
\end{table*}%

\begin{table*}
    \begin{tabular}{ccccccccc}
    \toprule
    \multicolumn{9}{c}{\textbf{wrong-ordering best-fit outcomes}}\\\cmidrule{2-7}
    \multirow{2}{*}{\textbf{parameter}} & \multicolumn{4}{c}{\textbf{true NO}} & \multicolumn{4}{c}{\textbf{true IO}} \\
    \cmidrule{3-4} \cmidrule{7-8}
     & inj. & JUNO & PINGU & comb. & inj. & JUNO & PINGU & comb. \\
    \midrule
    $\Delta m_{31}^{2}\;(\SI{e-3}{\electronvolt\squared}) $ & 2.525 & -2.496 & -2.386 & -2.490 & -2.438 & 2.466 & 2.565 & 2.472 \\
    $ \theta_{12}\;(\si{\degree})$ & 33.82 & 33.82 & - & 33.82 & 33.82 & 33.82 & - & 33.82 \\
    $ \theta_{13}\;(\si{\degree})$ & 8.61 & 8.63 & 8.71 & 8.76 & 8.65 & 8.60 & 8.57 & 8.47 \\
    $ \theta_{23}\;(\si{\degree})$ & 49.60 & - & 49.15 & 49.31 & 49.80 & - & 40.46 & 40.30$^\dagger$ \\
    \bottomrule
    \end{tabular}%
   \caption{Best-fit values within the wrong NMO for the free oscillation parameters after 6 years of operation, when fitting each experiment individually as well as for the combined fit (``comb.''). For a given true NMO, the column ``inj.'' specifies the injected parameter values, which are also given in Table~\ref{tab:usedParams}. The dagger denotes fit outcomes that correspond to a bound of a parameter's fit range.}
    \label{tab:paramFits}%
\end{table*}%

Table~\ref{tab:sens} gives the NMO sensitivity of the combined analysis for 6 years of data taking with each of the experiments, for both true NO and true IO using the nominal oscillation input models of Table~\ref{tab:usedParams}. As before, we assume the experiments' simultaneous start. The table also shows the experiments' stand-alone sensitivities as well as the sensitivities which each would obtain using the best-fit parameter values of the combined analysis, but not including prior penalties (labeled as ``statistical contribution''). On the one hand, this illustrates how the combined analysis profits simply from the tighter priors provided by the other experiment. On the other hand, the statistical contribution reveals the contribution of the synergetic effect to the sensitivity gain, as described in Sec.~\ref{sec:synergy}. In addition to the combination of JUNO and PINGU, which is subject to the strong synergy illustrated in the previous discussion, we here also evaluate the corresponding sensitivities based on the smaller-scope experimental setups: JUNO with a reduced number of eight reactor cores and the IceCube Upgrade.

\begin{figure*}
\centering
\input{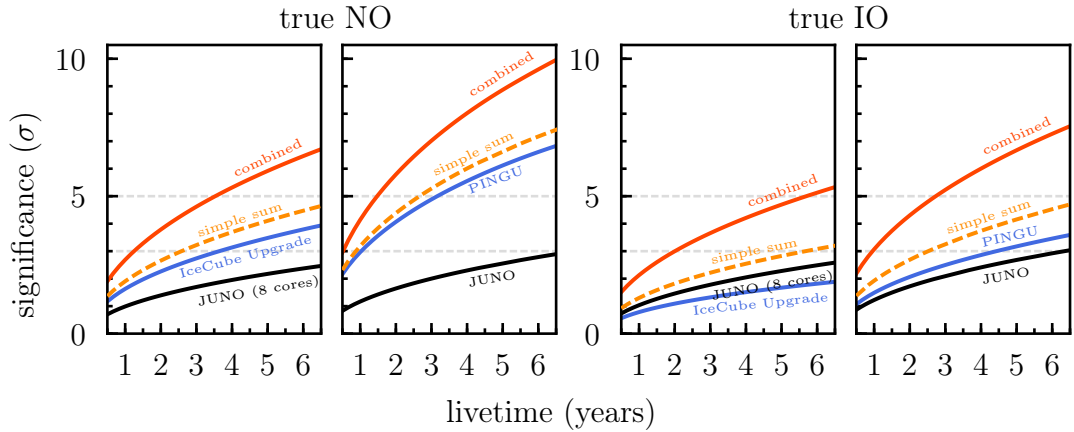}
\caption{\label{fig:livetimes}Livetime evolution of the NMO sensitivity of each considered pair of experiments: stand-alone, the simple (quadratic) sum, and the combination. Results for the nominal JUNO configuration and PINGU are shown side-by-side with the 8-core JUNO configuration (labeled as ``JUNO (8 cores)'') and the IceCube Upgrade. The two panels on the left assume true NO, while the two panels on the right assume true IO.}
\end{figure*}

Considering the stand-alone 8-core JUNO configuration, the sensitivity is projected to be around $2.5\sigma$ for either NMO, while that of the Upgrade is strongly dependent on which of the two input models is assumed to be true: for true NO, we obtain a significance of $3.8\sigma$, whereas for true IO we find $1.8\sigma$. The combined result exceeds the $5\sigma$ threshold considered as decisive. 

A similar picture emerges when one considers the nominal JUNO setup and PINGU, albeit at a higher significance-level: while we expect JUNO's stand-alone sensitivity to be close to $3\sigma$ for either NMO, the PINGU sensitivity in the case of true NO reaches more than $6\sigma$, and $3.5\sigma$ for true IO. The combined sensitivity is such that the NMO is established at exceedingly high levels of confidence of $9.6\sigma$ for true NO and $7.3\sigma$ for true IO.

Comparing the combined significances to the statistical contributions makes evident that the Upgrade and PINGU benefit far more from the oscillation parameter constraints provided by JUNO (in its reduced or nominal source configuration) than vice versa. As pointed out before, the main reason is found in the lower panels of Fig.~\ref{fig:synergy}, where the JUNO $\Delta m^2_{31}$ constraint is stronger than that provided by PINGU. As a result, the combined minimal $ \overline{\Delta\chi^2} $ lies close to the position of the minimum preferred by JUNO, creating a stronger tension with the data obtained by PINGU. The same holds true if one considers the 8-core JUNO configuration and the Upgrade instead.

The above effect is also illustrated in Table~\ref{tab:paramFits}, which lists the fit outcomes within the wrong ordering for the two stand-alone analyses of JUNO and PINGU and their combined analysis.
As expected, JUNO dominates the best-fit values of $ |\Delta m_{31}^{2}| $ and $ \theta_{12} $, whereas PINGU dominates the outcome in $ \theta_{23} $.

The combined best fit for $\theta_{13}$ (subject to the prior) prefers a value appreciably outside the range delimited by the outcomes of the two individual fits. A similar behavior could lead to a shift of the $ \theta_{13} $ best-fit value with respect to the truth when assuming the wrong ordering in a global fit.

\subsection{Livetime evolution of sensitivity}
Based on the assumption of a simultaneous start of data taking, Fig.~\ref{fig:livetimes} demonstrates the temporal evolution of the NMO sensitivity for a span between 1 year and 6 years of detector livetime, at the end of which the sensitivities reported in Sec.~\ref{sec:6years} are evaluated. The individual experiments' sensitivities are shown together with their simple (quadratic) sum and the combined sensitivities; true NO is depicted in the two leftmost panels, true IO in the two rightmost ones. Again, we investigate the joint analysis of the 8-core JUNO setup and the Upgrade, as well as the JUNO baseline together with PINGU.

One can see that the combination of the 8-core JUNO configuration and the Upgrade reaches $5\sigma$ in less than 4 years (true NO) respectively 6 years (true IO). The combined analysis of JUNO and PINGU reaches $5\sigma$ within about 1.5 years for true NO and within about 3 years of livetime of both detectors for true IO. Crucially, the comparison between the combined sensitivity and that resulting from the sum of each pair of experiments' $\overline{\Delta \chi^2}$ minima reveals that the synergy effect increases over time---the reason being that the minima in $ \Delta m_{31}^{2} $ within the wrong ordering become sharper. As a result, the minimal $\overline{\Delta\chi^2}$ grows faster when a full combined analysis is performed. In conclusion, the combined analysis does not only profit from the statistics gain over time but also from an enhancement of the synergy effect itself.

\begin{figure*}
\centering
\input{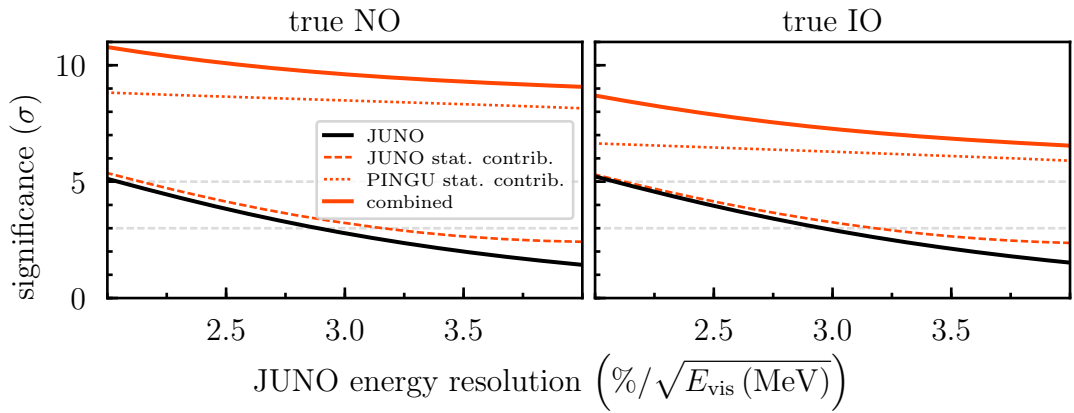}
\caption{\label{fig:eres}NMO sensitivities (combined, statistical contributions of JUNO and PINGU, JUNO stand-alone) as a function of JUNO's true energy resolution (for true NO on the left, true IO on the right) after 6 years of operation of both experiments.}
\end{figure*}

\begin{figure*}
\centering
\input{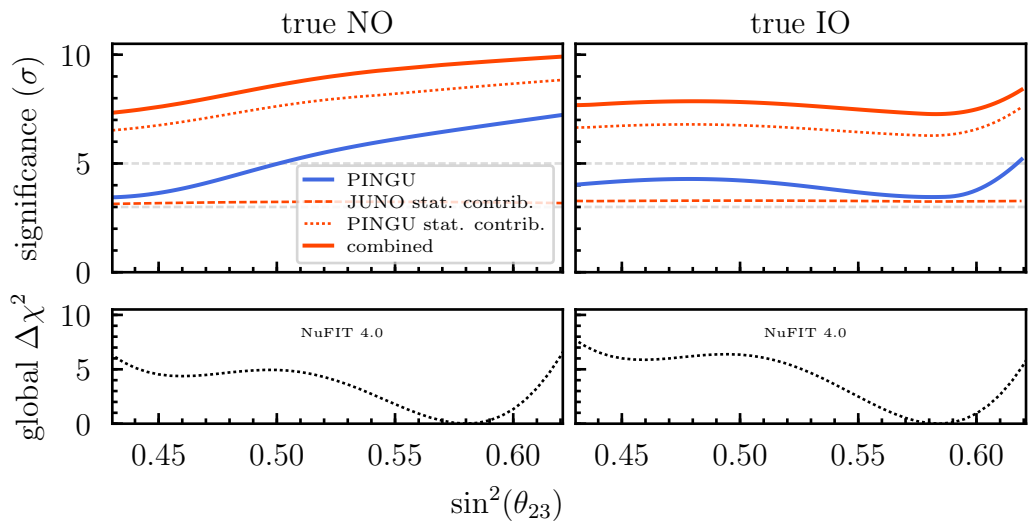}
\caption[True $ \theta_{23} $ dependency]{\label{fig:theta23}NMO sensitivities (combined, statistical contributions of JUNO and PINGU, PINGU stand-alone) as a function of the true value of $\sin^2(\theta_{23})$ (for true NO on the left, true IO on the right) after 6 years of operation of both experiments. The lower panels show the global $\Delta\chi^2$ constraint on $\sin^2(\theta_{23})$ (relative to the $\chi^2$ minimum within each ordering) from~\cite{Nufit4.0,nufit4.0url}.}
\end{figure*}

\subsection{Sensitivity dependence on true parameter values}
\label{subsec: dominant params}
In the following, we discuss the expected impact of the most important truth assumptions in the measurement of the NMO with JUNO and IceCube, both for the stand-alone experiments and in the context of the combined analysis. We focus on the dominant parameter for each experiment, namely the energy resolution in the case of JUNO and the atmospheric mixing angle $ \theta_{23} $ in IceCube. Similar to Sec.~\ref{sec:synergy}, we show our one-dimensional results for the baseline JUNO configuration and PINGU for ease of comparison with existing literature; the same qualitative behavior is observed for the 8-core JUNO setup and the Upgrade (Sec.~\ref{subsec:conservative scenario} provides the full two-dimensional study for that combination).

\subsubsection{JUNO energy resolution}
The detector's energy resolution (cf. Eq.~(\ref{eq:eres})) is the most critical parameter in the NMO measurement with JUNO. It is crucial for distinguishing between the rapid small-amplitude variations of the $\bar{\nu}_e$ energy spectrum brought about by the interference of the $\Delta m^2_{31}$- and $\Delta m^2_{32}$-driven oscillation modes in each of the possible NMO realizations.

The statement above is emphasized in Fig.~\ref{fig:eres}, which shows the dependence of the 6-year sensitivity on the energy resolution for JUNO alone as well as for the different types of combined analyses considered throughout this work. As expected, JUNO's sensitivity decreases as the energy resolution worsens. The projected median significance ranges from below $2\sigma$ to just above $5\sigma$ for the considered range in energy resolution (\SIrange{2}{4}{\%} at a visible energy of \SI{1}{\mega\electronvolt}). The fact that the PINGU sensitivity within the combined analysis decreases with worse energy resolution in JUNO arises due to the weakening JUNO constraint on $\Delta m^2_{31}$ within the wrong NMO. This effect is less prominent in the full combined analysis, however: JUNO profits more strongly from the combined analysis when its energy resolution worsens, as indicated by the separation between the solid black and dashed red lines. As the JUNO constraint on $\Delta m^2_{31}$ gains precision with improving energy resolution, the best-fit value of $ \Delta m_{31}^{2} $ for the combined analysis moves toward that preferred by JUNO, resulting in JUNO benefiting less from the combination.

The main takeaway, however, is that the combined sensitivity is rather stable with respect to the JUNO energy resolution: when a joint fit is performed, the significance lies well above the $5\sigma$ threshold even for the worst energy resolutions tested here.

Note that residual uncertainties in the calibration of the energy scale of the JUNO detector can lead to a reduced NMO sensitivity. As demonstrated in~\cite{Li:2013zyd}, this potential ambiguity can be effectively reduced by a self-calibration based on the oscillation pattern observed in the reactor $\bar{\nu}_e$ spectrum. However, even without self-calibration and for a scenario in which the wrong NMO becomes preferred over the true NMO, we have verified that the combined analysis of JUNO and PINGU delivers a high-significance rejection of the wrong NMO.

\subsubsection{Mixing angle $ \theta_{23} $}
Driving the overall strength of the observed oscillation signal, $ \theta_{23} $ is the dominant parameter regarding IceCube's NMO sensitivity. Figure~\ref{fig:theta23} shows the dependence of the 6-year sensitivity on $\sin^2(\theta_{23})$ for PINGU alone, for the combined analysis of both experiments, and for their respective contributions to the combined sensitivity. JUNO's stand-alone NMO sensitivity is not shown since its event spectrum is unaffected by $ \theta_{23} $.

For the considered range in $\sin^2(\theta_{23})$, PINGU's projected median significance ranges from around $3\sigma$ to $7\sigma$ when NO is true, and from around $3\sigma$ to $5\sigma$ when IO is true. The behavior of PINGU within the combined analysis is nearly the same as in the stand-alone case, with the NMO sensitivity being shifted to a higher value due to the synergy effect. This is not obvious because PINGU is not free to choose the values for $ \Delta m_{31}^{2} $ and $\theta_{13}$ anymore, which are dominated ($ \Delta m_{31}^{2} $) respectively affected ($ \theta_{13} $) by JUNO. Also within the combined analysis, JUNO's contribution to the NMO sensitivity is only barely affected by the true value of $ \sin^2(\theta_{23}) $. This result is not necessarily obvious either because PINGU's dependence on $ \sin^2(\theta_{23}) $ could lead to shifts in the best-fit values for other oscillation parameters and therefore indirectly affect the JUNO result as well.

As a result of the considerations above, the behavior of the projected combined sensitivity is similar to PINGU's, though somewhat more stable with respect to the true value of $\sin^2(\theta_{23})$. Again, when a joint fit is performed, the significance lies well above the $5\sigma$ threshold even for the least favorable values of $\sin^2(\theta_{23})$ tested here.

\subsection{NMO potential for JUNO with eight cores and the Upgrade}
\label{subsec:conservative scenario}
Assuming $5\sigma $ as the target sensitivity, neither the Upgrade, nor the 8-core JUNO configuration, nor the simple sum of their stand-alone sensitivities is expected to lead to a decisive, $>5\sigma$, determination of the NMO  ($\sim$ 5 years of joint operation). However, the boost in sensitivity due to a combined fit is so substantial that this target is attainable, cf. Fig.~\ref{fig:livetimes} for our nominal truth assumptions about the JUNO energy resolution and $\sin^2(\theta_{23})$.

\begin{figure*}[th]
\centering
\input{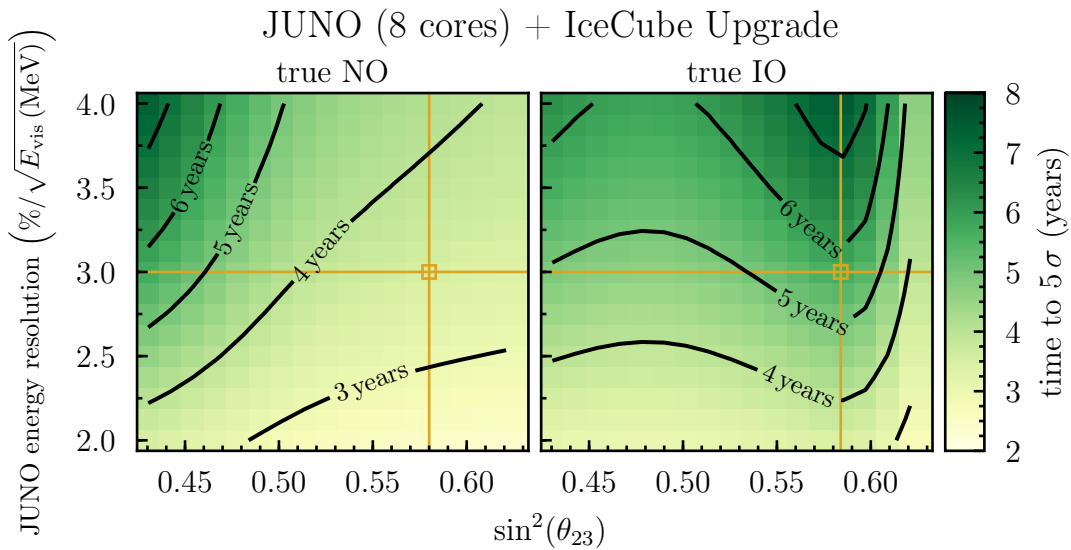}
\caption{Time required for the combined analysis of the 8-core JUNO configuration and the IceCube Upgrade to attain a $5\sigma$ measurement of the NMO as a function of the true mixing angle $\sin^2\left(\theta_{23}\right)$ and JUNO's energy resolution, for true NO on the left and true IO on the right. The empty square marks our nominal assumption about the two parameters. Solid contours trace parameter combinations for which the required time is 3, 4, 5, 6, and 7 years, respectively.}
\label{fig:time_to_5}
\end{figure*}

Going beyond the nominal scenario for this pair of parameters, Fig.~\ref{fig:time_to_5} explores the corresponding two-dimensional parameter space and shows the time needed to obtain a significance of $5\sigma$ in a combined analysis of the 8-core JUNO configuration and the Upgrade at each point. Here, the nominal values are marked by the two orthogonal lines, and the nominal point by the empty square. The dependencies roughly follow the behavior of the combined sensitivity of JUNO and PINGU shown as a function of the true value of each of the parameters separately in Fig.~\ref{fig:eres} and Fig.~\ref{fig:theta23}. For true NO on the left, the least favorable point is located in the upper left hand corner, where the true value of $\sin^2(\theta_{23})$ is smallest ($\sim 0.43$) and JUNO's energy resolution is worst ($\sim \SI{4}{\percent}$). Here, the NMO can only be determined after about 7 years of data taking. Conversely, when $\sin^2(\theta_{23}) \approx 0.63$ and the JUNO energy resolution amounts to $\SI{2}{\percent}$, the required time is reduced to around 2.5 years. For true IO on the right, the least favorable value of $\sin^2(\theta_{23})$ approximately coincides with our nominal assumption; depending on the JUNO energy scale, the time to determine the NMO ranges from below 4 years to more than 7 years. The most favorable scenario is again located in the lower right hand corner, where less than 3 years of measurement are expected to suffice in order to determine the NMO.

\section{Combined oscillation parameter sensitivities}
\label{chapter4}
The synergy effect on the NMO sensitivity discussed throughout the previous chapter could imply a similar gain in the measurement of the oscillation parameters to which both JUNO and IceCube are sensitive, i.e., $ \theta_{13} $ and $\Delta m^2_{31}$. In this chapter we again focus on a combined analysis of JUNO (with ten cores) and PINGU (instead of the Upgrade) to cover the most powerful scenario.

\begin{figure*}
\centering
\input{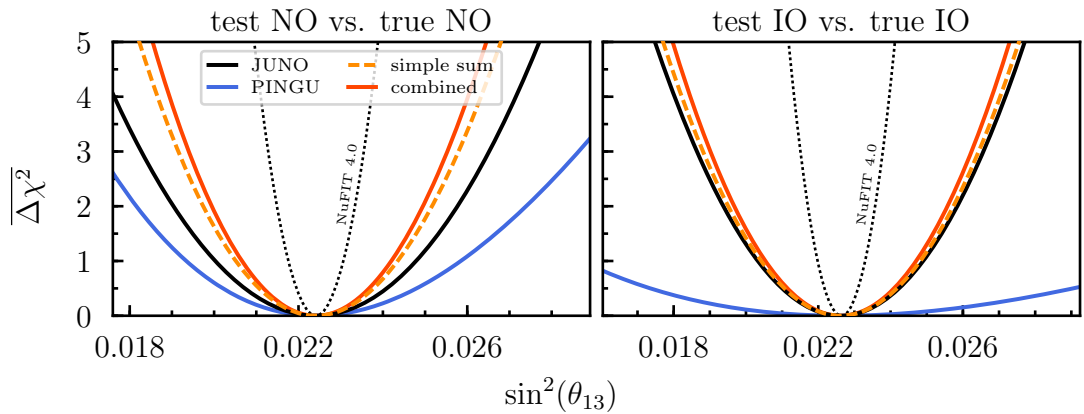}
\caption[$ \theta_{13} $ sensitivities]{$\overline{\Delta \chi^2}$ profiles as a function of the tested values of $\sin^2(\theta_{13})$ within the true ordering for JUNO and PINGU stand-alone, their simple sum, and their combination, for a livetime of 6 years of both experiments. On the left (right), the NO (IO) is assumed to be true. The current global sensitivity to $\sin^2(\theta_{13})$ is superimposed (dotted black)~\cite{Nufit4.0,nufit4.0url}.}
\label{fig:theta13_sens}
\end{figure*}

Assuming the NMO has been correctly identified, the two panels in Fig.~\ref{fig:theta13_sens} show both JUNO's and PINGU's as well as their combined sensitivity to $ \sin^2(\theta_{13}) $ in the absence of external constraints on this mixing angle, for true NO on the left and true IO on the right. In both cases, we superimpose the current global $\Delta \chi^2$ constraint~\cite{Nufit4.0,nufit4.0url} in the same plot. As expected, the combined analysis is able to measure $ \sin^2(\theta_{13}) $ with a slightly higher precision than that obtained via the simple sum of the two stand-alone profiles. It does not yield an improvement over current global constraints though.

\begin{figure*}
\centering
\input{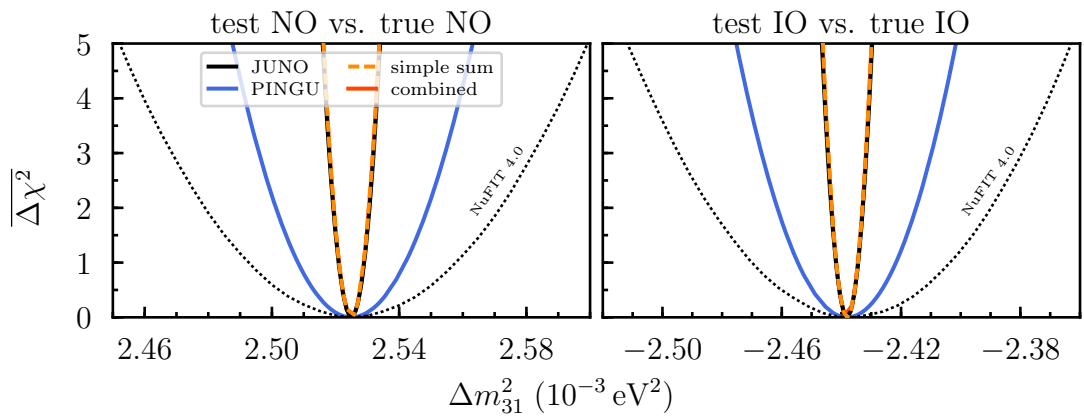}
\caption[$ \Delta m^2_{31} $ sensitivities]{$\overline{\Delta \chi^2}$ profiles as a function of the tested values of $\Delta m^2_{31}$ within the true ordering for JUNO and PINGU stand-alone, their simple sum, and their combination, for a livetime of 6 years of both experiments. On the left (right), the NO (IO) is assumed to be true. The current global sensitivity to $\Delta m^2_{31}$ is superimposed (dotted black)~\cite{Nufit4.0,nufit4.0url}.}
\label{fig:deltam31_sens}
\end{figure*}

Fig.~\ref{fig:deltam31_sens} shows the analogous information for $ \Delta m^2_{31} $. In this case, the stand-alone JUNO measurement outperforms the global constraints, and so does PINGU's measurement---albeit to a lesser extent. The combination of JUNO and PINGU, however, does not lead to any further gain in precision compared to that obtained by JUNO alone, no matter whether one takes the simple sum or performs a combined fit.

In summary, as opposed to the case of the NMO measurement, the combined analysis of JUNO and PINGU does not lead to a significant enhancement of the constraints on the oscillation parameters $\theta_{13}$ and $\Delta m^2_{31}$. In the former case the existing global constraints are stronger, whereas in the latter case the projected JUNO stand-alone sensitivity is the same as that of the combined fit. We have verified that these conclusions apply identically also to the combined analysis of JUNO with eight cores and the Upgrade.

\section{Conclusion}
\label{chapter5}
In this paper we investigate the potential of a combined neutrino mass ordering analysis of the data from the future reactor neutrino experiment JUNO and the future atmospheric neutrino experiments IceCube Upgrade and PINGU.

Owing to the different positions of the minima in the oscillation parameter space within the wrong-ordering hypothesis, the combined analysis of JUNO and the IceCube Upgrade or PINGU achieves an NMO sensitivity which exceeds the purely statistical combination of their stand-alone sensitivities. This synergy effect is based on the distinct experimental setups---for which the dominant neutrino oscillation channels differ, as well as neutrino energies, baselines, and the relevance of Earth matter effects---and is most prominent in $ \Delta m_{31}^{2} $. Both JUNO and PINGU will have sensitivity to this mass-squared difference much beyond that of the existing generation of oscillation experiments or their combination. In this regard, the IceCube Upgrade and PINGU benefit far more from the combined analysis than JUNO. The reason is that JUNO's precise $ \Delta m_{31}^{2} $ constraint within the wrong ordering is the actual driver of the synergy effect.

It should be stressed that the greatest sensitivity benefit is achieved via the combination of JUNO with an NMO-sensitive long-baseline experiment. The synergy here arises from the complementarity between the two measurement methods, i.e., sub-dominant vacuum oscillations versus long-baseline oscillations enhanced by matter effects. Both of these methods result in distinct oscillation patterns when comparing the two possible neutrino mass ordering realizations. Substantially smaller benefit is expected when including more long-baseline experiments.

Our studies demonstrate that the combined analysis of JUNO with eight reactor cores and the IceCube Upgrade is projected to result in a significance of $5\sigma$ within approximately 3 years to 7 years of livetime. 
In the most promising case, corresponding to the combined analysis of JUNO with its full reactor configuration and PINGU, an NMO significance of $5\sigma$ can be reached with less than 2 years of data taking. Thus, in brief, a combined analysis with JUNO and IceCube will determine the neutrino mass ordering at a significance beyond the $5\sigma$ level within the expected operation times of both experiments, even for a more conservative scenario and for unfavorable regions of parameter space.

Finally, we note that a combined measurement of the oscillation parameters of the PMNS paradigm~\cite{Giganti:2017fhf}, such as $\sin^2(\theta_{13})$ or $\Delta m^2_{31}$, does not significantly improve the stand-alone capabilities or the measurements with the existing generation of experiments.

\begin{acknowledgements}
The IceCube collaboration gratefully acknowledges the support from the following agencies and institutions: USA {\textendash} U.S. National Science Foundation-Office of Polar Programs,
U.S. National Science Foundation-Physics Division,
Wisconsin Alumni Research Foundation,
Center for High Throughput Computing (CHTC) at the University of Wisconsin-Madison,
Open Science Grid (OSG),
Extreme Science and Engineering Discovery Environment (XSEDE),
U.S. Department of Energy-National Energy Research Scientific Computing Center,
Particle astrophysics research computing center at the University of Maryland,
Institute for Cyber-Enabled Research at Michigan State University,
and Astroparticle physics computational facility at Marquette University;
Belgium {\textendash} Funds for Scientific Research (FRS-FNRS and FWO),
FWO Odysseus and Big Science programmes,
and Belgian Federal Science Policy Office (Belspo);
Germany {\textendash} Bundesministerium f{\"u}r Bildung und Forschung (BMBF),
Deutsche Forschungsgemeinschaft (DFG),
Helmholtz Alliance for Astroparticle Physics (HAP),
Initiative and Networking Fund of the Helmholtz Association,
Deutsches Elektronen Synchrotron (DESY),
and High Performance Computing cluster of the RWTH Aachen;
Sweden {\textendash} Swedish Research Council,
Swedish Polar Research Secretariat,
Swedish National Infrastructure for Computing (SNIC),
and Knut and Alice Wallenberg Foundation;
Australia {\textendash} Australian Research Council;
Canada {\textendash} Natural Sciences and Engineering Research Council of Canada,
Calcul Qu{\'e}bec, Compute Ontario, Canada Foundation for Innovation, WestGrid, and Compute Canada;
Denmark {\textendash} Villum Fonden, Danish National Research Foundation (DNRF), Carlsberg Foundation;
New Zealand {\textendash} Marsden Fund;
Japan {\textendash} Japan Society for Promotion of Science (JSPS)
and Institute for Global Prominent Research (IGPR) of Chiba University;
Korea {\textendash} National Research Foundation of Korea (NRF);
Switzerland {\textendash} Swiss National Science Foundation (SNSF);
United Kingdom {\textendash} Department of Physics, University of Oxford.

This work has been supported by the Cluster of Excellence ``Precision Physics, Fundamental Interactions, and Structure of Matter'' (PRISMA+ EXC 2118/1) funded by the German Research Foundation (DFG) within the German Excellence Strategy (Project ID 39083149).
\end{acknowledgements}

\appendix


\bibliographystyle{elsarticle-num}
\bibliography{textbib}

\end{document}